\documentclass[aps,prb,twocolumn,superscriptaddress,showpacs]{revtex4}
\usepackage{graphicx}
\usepackage{latexsym}
\usepackage{amssymb}
\usepackage{amsmath}
\usepackage{amsfonts}
\usepackage{bm}
\usepackage{multirow}
\usepackage[usenames,dvipsnames]{xcolor}
\newcommand{\bra}[1]{\langle #1|}
\newcommand{\ket}[1]{|#1 \rangle}
\newcommand{\dd}{\mathrm{d}}
\newcommand{\ii}{\mathrm{i}}
\newcommand{\dsZ}{\mathbb{Z}}
\newcommand{\Tr}{\operatorname{Tr}}

\newcommand{\vect}[1]{{\bm{#1}}}
\newcommand{\gate}[1]{{\textsf{#1}}}
\newcommand{\norm}[1]{{\lVert #1\rVert}}
\newcommand{\hl}[1]{{\color{BrickRed}\bf{#1}}}
\newcommand{\beq}{\begin{equation}}
\newcommand{\eeq}{\end{equation}}
\newcommand{\eqnref}[1]{Eq.\,\eqref{#1}}
\newcommand{\figref}[1]{Fig.\,\ref{#1}}
\newcommand{\tabref}[1]{Tab.\,\ref{#1}}

\begin{document}

\title{Entanglement Holographic Mapping of Many-Body Localized System\\ by Spectrum Bifurcation Renormalization Group}
\author{Yi-Zhuang You}
\affiliation{Department of physics, University of California,
Santa Barbara, CA 93106, USA}

\author{Xiao-Liang Qi}
\affiliation{Department of physics, Stanford University, CA 94305, USA}

\author{Cenke Xu}
\affiliation{Department of physics, University of California,
Santa Barbara, CA 93106, USA}

\date{\today}

\begin{abstract}

We introduce the spectrum bifurcation renormalization group (SBRG) as an improved application of the excited-state real space renormalization group (RSRG-X) for a class of qubit models. Starting from a disordered many-body Hamiltonian in the full many-body localized (MBL) phase, SBRG flows to the MBL fixed-point Hamiltonian, and generates the local integrals of motion and the matrix product state representations for all eigenstates. The method is applicable to many interacting spin and fermion models in the full MBL phase. As a Hilbert-space preserving RG, SBRG also generates an entanglement holographic mapping, which duals the MBL state to a fragmented holographic space.

\end{abstract}


\maketitle

\tableofcontents

\section{Introduction}

The phenomenon of many-body localization (MBL)\cite{Anderson:1980,BAA:2006,GMP:2005,Imbrie:2014,Huse:2015rv} has attracted much research interest recently.\cite{Znidaric:2008he,Huse:2007is,Huse:2010he,Monthus:2010fm,Berkelbach:2010ir,Moore:2012ge,Abanin:2013ge,Huse:2013ci,Vishwanath:2013tp,Bauer:2013al,Swingle:2013oy,Altman:2013rg,Refael:2013du,Altman:2014dq,Altman:2014hg,Abanin:2013lc,Huse:2014ec,Abanin:2015io,Andraschko:2014vz,Reichman:2014hz,Pollmann:2014sf,Chandran:2014fv,Kim:2014zj,Chandran:2014tn,Abanin:2014qq,Moore:2015qr,Reichman:2015sf,Bloch:2015ca,Chandran:2015rc,Vishwanath:2015po,Xu:2015gg} It is the generalization of the Anderson localization\cite{Anderson:1958} to interacting systems, and can be described as the Fock space localization\cite{Avishai:1996qe,Avishai:1997cs} of finite energy density states in the presence of strong disorder. An MBL system cannot serve as its own heat bath, and thus violates the eigenstate thermalization hypothesis (ETH).\cite{Deutsch:1991ik,Srednicki:1994ns,Tasaki:1998vy} From many aspects, the MBL excited states are like ground states.\cite{Huse:2013ci,Vishwanath:2013tp}  For example, the entanglement entropy (EE)\cite{Horodecki:2009gf} follows the area-law scaling\cite{Srednicki:1993rv} in the MBL state,\cite{Bauer:2013al,Swingle:2013oy} in contrast to the volume-law scaling\cite{Page:1993fv,Foong:1994bf,Sen:1996rw} in the ETH state. In the full MBL system,\cite{Huse:2014ec} all energy eigenstates are localized and have area-law entanglement entropy, which implies the existence of matrix product state (MPS)\cite{Hastings:2007sd,Verstraete:2006qt,Eisert:2010rz} representations for all eigenstates.

A natural question is whether there is an efficient way to find (or approximately find) the MPS representation for every eigenstate in the full many-body spectrum of a MBL system? In fact, there have been several nice approaches trying to answer this question, either by matching the unitary circuit of exact diagonalization,\cite{Clark:2014ed} or by the spectral tensor network for local integrals of motion,\cite{Chandran:2014tn,Kim:2014zj} or by variational tensor network optimization.\cite{Pollmann:2015vn} Here we would like to tackle the problem from a different angle using the renormalization group (RG) approach. For example, the density matrix renormalization group (DMRG)\cite{White:1992lo,Schollwock:2011ud} has been shown to be a powerful and successful method to find the MPS representations for ground states. For full MBL systems, DMRG can be generalized to target highly-excited states in the many-body spectrum as well, known as DMRG-X.\cite{Khemani:2015qf,Yu:2015vn,Lim:2015zr,Kennes:2015yg} DMRG-X is a non-perturbative high-accuracy approach to find the MPS representation for an individual excited state at a given energy-density. However it is still quite expansive to obtain all the eigenstates using DMRG-X. We would like to propose a less accurate but more efficient RG approach to resolve all eigenstates in the MBL spectrum simultaneously. Obviously, we must take advantage of the strong disorder nature of the MBL system in order to target the full spectrum. The real-space renormalization group (RSRG)\cite{Dasgupta:1980so,Lee:1982jv,Fisher:1992is,Fisher:1994he,Fisher:1995cr,Huse:2001ez} (also known as the strong disorder renormalization group) is such an RG scheme that is designed to work with strong disorder. The original proposal of RSRG is ground state targeting, and it was recently extended to target all excited states as RSRG-X.\cite{Altman:2013rg,Altman:2014dq,Altman:2014hg,Refael:2013du,Potter:2015th} RSRG-X finds all eigenstates by transversing the spectrum branching tree explicitly. However for a specific class of qubit models (such as quantum Ising models), where the spectrum branches at each RG step to two subspaces with identical dimensions, it is then possible to implicitly encode the spectrum branching by an emergent qubit degree of freedom, and formulate the RG as a flow of the entire many-body Hamiltonian without branching to any specific energy subspaces. 

The idea of formulating RSRG-X as a Hamiltonian flow has appeared in several recent works Ref.\,\onlinecite{Altman:2013rg,Swingle:2013oy,Refael:2013du}. In this work, we will introduce a systematic RG scheme along this line of thought, as outlined in the following.  First we pick out the leading energy scale (highest frequency, largest coupling) term $H_0$ in the Hamiltonian $H$, and rotate the Hamiltonian $H$ to the diagonal basis of $H_0$ by a local unitary transformation. Next we eliminate the terms that anti-commute with the leading term $H_0$ by second order perturbation, and obtain a new Hamiltonian. Then we turn to the next leading energy scale term in the new Hamiltonian and repeat the above RG procedure, until all degrees of freedom in the system are accounted for. Finally, we collect the unitary transformations that have been performed along the RG flow, and arrange them into a tensor network, then the MPS representations for all eigenstates are found and encoded in the tensor network. Because each RG step bifurcates the spectrum until the Hamiltonian is eventually diagonalized, we decided to name the RG scheme as the \emph{spectrum bifurcation} renormalization group (SBRG).

SBRG is a Hilbert space-preserving RG scheme, keeping track of the flow of the whole many-body Hamiltonian without truncating the Hilbert space to any specific branch of the spectrum. It can be considered as an improvement of RSRG-X, in the sense that SBRG allows generation of new terms under the Hamiltonian flow, and also the history dependence of the spectrum branching is not explicitly realized but implicitly encoded in the ultraviolet-infrared (UV-IR) mixing terms in the Hamiltonian. Therefore SBRG can be applied to strong interacting models beyond the limitation of a closed-form RG.  Like RSRG-X, SBRG also targets the full many-body spectrum rather than just the ground state, and focuses on small frequency (small energy difference) rather than the low absolute energy. However, SBRG also suffers from the same limitation as RSRG that it is a basis dependent approach, which means local unitary rotations of the Pauli basis will affect the RG flow.\footnote{We thank the referee for pointing out the basis dependence problem to us.} This limitation may be overcome by resolving the local basis rotation before identifying the leading energy scale, however we will leave the possible improvements for future study.

Another motivation comes from the recent research effort to interpret the RG transformation as a realization of the holographic duality,\cite{Swingle:2012bs,Swingle:2012yq,Balasubramanian:2013xz,Leigh:2014rb,Lee:2014qh} examples include the multiscale entanglement
renormalization ansatz (MERA) network\cite{Vidal:2008zp,Takayanagi:2015so,Czech:2015ig,Bao:2015hf}, cMERA\cite{Verstraete:2013do,Takayanagi:2012vm,Takayanagi:2014an,Takayanagi:2015pd,Molina-Vilaplana:2015lr}, ab initio holography\cite{Lee:2015aa} ect. In particular, Ref.\,\onlinecite{Qi:2013fm,Qi:2015ct} proposed the entanglement\footnote{Originally EHM is short for ``exact'' holographic mapping. However, in this paper, EHM circuit is actually an approximation of the exact diagonalization circuit, so to avoid confusion of the mean of ``exact'', we change the name to ``entanglement'' holographic mapping.} holographic mapping (EHM) as  a unitary mapping between boundary (physical) degrees of freedom and bulk (emergent) degrees of freedom for every Hilbert space-preserving RG. Here we apply the same idea to SBRG for the MBL system, and it turns out that the emergent degrees of freedom are just the emergent conserved quantities\cite{Abanin:2013lc,Huse:2014ec,Abanin:2015io} in the MBL fixed-point Hamiltonian.

The emergent conserved quantities are identified by SBRG progressively as the leading energy scale. The spectrum bifurcation at each RG step is controlled by the emergent conserved quantities (like controlled-gates in the quantum circuit). In the end of the RG flow, the MBL fixed-point Hamiltonian will emerge. By collecting the unitary transformations that have been performed to the Hamiltonian during the RG flow as matrix product operators (MPO),\cite{Verstraete:2004xj} one can reconstruct the approximate tensor network representation of the full many-body spectrum, which encodes the approximate MPS representation of each MBL eigenstate. An important observation of this work is that the EHM circuit of SBRG can be approximated by a Clifford circuit\cite{Gottesman:1998lk,Gottesman:1998cl} which enables efficient calculation of many physical properties of the MBL system.

The paper is organized as follows. In Section \ref{sec: SBRG}, we start by introducing SBRG algorithm under generic settings, and discuss the limitations and strengths of the method. In Section \ref{sec: application}, we apply SBRG to the disordered quantum Ising model, and benchmark the energy spectrum and eigenstates obtained by SBRG. We will demonstrate that SBRG flows towards the strong disorder limit in the MBL phase. The RG flow also generates a Clifford circuit which encodes the (approximate) MPS representations of all eigenstates. In Section \ref{sec: EHM}, we interpret the Clifford circuit as an EHM tensor network. We investigate the properties of the emergent conserved quantities and confirm the area-law entanglement in the MBL phase. Finally we will discuss the locality in the holographic bulk, and the collapse of SBRG near thermalization.

\section{Spectrum Bifurcation Renormalization Group}\label{sec: SBRG}

\subsection{SBRG Algorithm}

The basic idea of SBRG is to progressively identify emergent conserved quantities by block-diagonalization of the leading energy scale, and eliminate the block-off-diagonal terms in the Hamiltonian by second order perturbation. The most general setting is to start from the following qubit model Hamiltonian
\beq\label{eq: H qubit model}
H=\sum_{[\mu]}h_{[\mu]}\sigma^{[\mu]}=\sum_{[\mu]}h_{[\mu]}\underset{i}{\otimes}\sigma^{\mu_i},
\eeq
where $\sigma^{\mu_i}$ ($\mu_i = 0,1,2,3$) denotes the Pauli matrix acting on the $i$th qubit. $\sigma^{[\mu]}\equiv\sigma^{\mu_1\mu_2\mu_3\cdots}\equiv \sigma^{\mu_1}\otimes\sigma^{\mu_2}\otimes\sigma^{\mu_3}\otimes\cdots$ is a short-handed notation for the direct product of Pauli matrices, or called \emph{Pauli operators}. The Hamiltonian is simply a sum of the Pauli operators $\sigma^{[\mu]}$ with real coefficients $h_{[\mu]}$ over all Pauli indices $[\mu]$, which are randomly drawn from independent distributions. Here we have assumed the Hilbert space dimension is a power of two as $2^N$, such that any Hamiltonian in the Hilbert space can be written as a linear combination of the Pauli operators $\sigma^{[\mu]}$. The qubit model is quite general. It allows multi-qubit non-local interactions on a generic lattice, and can describe a large class of spin and fermion systems in all dimensions (fermion models can be mapped to qubit models by Jordan-Wigner transformation\footnote{In higher spacial dimension, the fermion model will be mapped to the non-local qubit model in general. Although SBRG algorithm complexity may grow exponentially in the absence of locality, we may still consider the non-local qubit model as our most generic stating point for theoretical discussions.}).

Starting from the qubit model \eqnref{eq: H qubit model}, SBRG algorithm goes as follows. First, we pick out the leading energy scale term in the Hamiltonian, which amounts to selecting the term $h_{[\mu]}\sigma^{[\mu]}$ with the maximal coefficient $|h_{[\mu]}|$ (in its absolute value) among all terms in the Hamiltonian $H$, and denote it as
\beq H_0=h_3\sigma^{[\mu]_\text{max}}, \eeq
where $|h_3|\equiv|h_{[\mu]_\text{max}}|=\max_{[\mu]}|h_{[\mu]}|$ represents the leading coefficient (the meaning of the subscript 3 will be evident later). Because each Pauli operator $\sigma^{[\mu]}$ has normalized eigenvalues $\pm1$, the energy scale of each term is only determined by the coefficient $h_{[\mu]}$ in the front. At this point, we need to assume that there is only one unique term with the leading energy scale, and all the other terms have energy scales sufficiently less than the leading one. This assumption can be justified in the strong disorder limit, but will break down for uniform (translational invariant) systems.

Then we block-diagonalize the leading term $\sigma^{[\mu]_\text{max}}$ by a Clifford group rotation $R$ such that
\beq
\sigma^{[\mu]_\text{max}}\to R^\dagger \sigma^{[\mu]_\text{max}}R = \sigma^{3[0\cdots]},
\eeq
and at the same time transform all the other terms in the Hamiltonian by the same rotation, such that $H\to R^\dagger H R$. Here $\sigma^{3[0\cdots]}$ denotes $\sigma^3$ times the rest of the identity matrices. As a technical note, although we write the diagonal form as $\sigma^{3[0\cdots]}$ for theoretical formulation of the RG scheme, in practice the $\sigma^3$ operator does not need to be swap to the first qubit literally. Its action can remain in the local support of the original Pauli operator $\sigma^{[\mu]_\text{max}}$, such that $R$ is a local unitary transformation (see \figref{fig: RG step} later for an explicit illustration, and also Appendix \ref{sec: R transform} for implementation details). The key observation is that $R$ is not a generic unitary transformation, but an element in the Clifford group which rotates one Pauli operator to another. So $R$ can be easily found for any given $\sigma^{[\mu]_\text{max}}$ (see Appendix \ref{sec: R transform} for the algorithm), and  can be applied to the Hamiltonian efficiently. This is related to the Gottesman-Knill theorem\cite{Gottesman:1998cl} that Clifford circuits can be simulated efficiently on a classical computer. 

As we block-diagonalize the leading energy scale, $H_0$ becomes $H_0=h_3\sigma^{3[0\cdots]}$, and the many-body spectrum bifurcates to the high-energy $E\simeq |h_3|$ and the low-energy $E\simeq-|h_3|$ sectors (blocks). We must reduce other terms in the Hamiltonian into either the higher or the lower energy sectors. Thus we classify the terms by their commutativity with $H_0$ as
\beq\label{eq: H in RG}
H = H_0 + \Delta + \Sigma,
\eeq
where $\Delta$ are terms that commute with $H_0$, i.e. $\Delta H_0 = H_0 \Delta$; and $\Sigma$ are terms that anti-commute with $H_0$, i.e. $\Sigma H_0 = -H_0 \Sigma$. All terms must fall into these two classes, because $H_0$ contains only one single term of a Pauli operator (not the sum of several terms). Given $H_0=h_3\sigma^{3[0\cdots]}$ in the $R$-rotated basis, one can see that $\Delta$ rests in the diagonal block as a combination of $\sigma^{0[\cdots]}$ and $\sigma^{3[\cdots]}$ , and $\Sigma$ rests in the off-diagonal block as a combination of $\sigma^{1[\cdots]}$ and $\sigma^{2[\cdots]}$. The diagonal terms $\Delta$  is left untouched, and are passed down with the Hamiltonian to the next step of SBRG. The off-diagonal terms $\Sigma$ must be renormalized by 2nd order perturbation, which corresponds to the unitary transformation $H\to S^\dagger HS$ (known as the Schrieffer-Wolff transformation\cite{Schrieffer:1966nu}) with
\beq\label{eq: SW transform}
S=\exp\Big(-\frac{1}{2h_3^2}H_0\Sigma\Big),
\eeq
carried out to the 2nd order of $h_3^{-1}$ (see Appendix \ref{sec: S transform} for derivation). The resulting effective Hamiltonian within the diagonal blocks reads 
\beq\label{eq: H 2nd order}
\begin{split}
H&=H_0+\Delta - \frac{1}{2} \Sigma H_0^{-1}\Sigma\\
&=H_0+\Delta+\frac{1}{2h_3^2} H_0\Sigma^2.
\end{split}
\eeq
Under the Schrieffer-Wolff transformation $S$, the Hamiltonian is block-diagonalized (to the 2nd order), as one can check that $H_0\Sigma^2$ now commutes with $H_0$. Since new terms are generated under the 2nd order perturbation, the number of terms in the Hamiltonian will presumably grow in this step. In practice, small terms can be truncated to control the growth rate (see Appendix \ref{sec: S transform} for the truncation scheme).

So finally, the new Hamiltonian takes the form of
\beq
H=h_{3}\sigma^{3[0\cdots]}+\sum_{\lambda,[\mu]}h_{\lambda[\mu]}\sigma^{\lambda[\mu]},
\eeq
where the first qubit is acted by $\sigma^\lambda$ with $\lambda=0$ or 3 only, and the remaining qubits are acted by the generic $\sigma^\mu$ with $\mu=0,1,2,3$. The leading term $h_{3}\sigma^{3[0\cdots]}$ is singled out\footnote{The coefficient $h_3$ here is slightly different from the $h_3$ used previously, in that the 2nd order correction to the $\sigma^{3[0\cdots]}$ term has been absorbed into the new $h_3$ coefficient here.}  and is then ascribed to the effective Hamiltonian (which will not be touched in the later RG steps). The operator $\sigma^{3[0\cdots]}$ is also identified as an \emph{emergent conserved quantity}, because its energy scale is so high (compared to its local neighbors) that it is unlikely to flip as we zoom in the spectrum towards the low frequency limit. The remaining terms $H_\text{res}=\sum_{\lambda[\mu]}h_{\lambda[\mu]}\sigma^{\lambda[\mu]}$ have the same form as the qubit model \eqnref{eq: H qubit model} that we started with. They will enter the next SBRG step. In the next step, we continue to pick out the leading energy scale term in $H_\text{res}$, diagonalize it by a Clifford group rotation to the second qubit as $\sigma^{03[0\cdots]}$, and reduce the off-diagonal terms by the 2nd order perturbation. Because $H_\text{res}$ commutes with $\sigma^{3[0\cdots]}$, it can be diagonalized by the $\sigma^{3[0\cdots]}$-preserving unitary transformations, meaning that the previously identified conserved quantity $\sigma^{3[0\cdots]}$ will not be affected by the later RG transformations. 

As we keep running SBRG procedure, the Hamiltonian will flow to the following generic form, which splits into two parts (where $\lambda=0,3$ and $\mu=0,1,2,3$)
\beq
\begin{split}
H&=H_\text{eff}+H_\text{res},\\
H_\text{eff}&=\sum_{[\lambda]}h_{[\lambda]}\sigma^{[\lambda][0\cdots]},\\
H_\text{res}&=\sum_{[\lambda],[\mu]}h_{[\lambda][\mu]}\sigma^{[\lambda][\mu]}.
\end{split}
\eeq
The \emph{effective Hamiltonian} $H_\text{eff}$ contains all the terms that have been fully diagonalized in previous RG steps, and the \emph{residual Hamiltonian} $H_\text{res}$ contains the remaining terms to be diagonalized in future RG steps. The Hilbert space is also naturally partitioned into the \emph{emergent} Hilbert space acted by $\sigma^{[\lambda]}$ ($\lambda=0,3$), and the \emph{physical} Hilbert space acted by $\sigma^{[\mu]}$ ($\mu=0,1,2,3$). In each RG step, the emergent Hilbert space grows by one qubit, while the physical Hilbert space shrinks by one qubit. Correspondingly the terms in $H_\text{res}$ are progressively renormalized to $H_\text{eff}$. Each emergent qubit is an emergent conserved quantity identified at a particular energy scale, which controls the branching of the spectrum of that energy scale. Since the Hilbert space is not truncated, the information of spectrum bifurcation is kept in the $[\lambda]$ dependence of the $\sigma^{[\lambda][\mu]}=\sigma^{[\lambda]}\otimes\sigma^{[\mu]}$ terms in $H_\text{res}$. If we specify an Ising configuration $\ket{\tau}$ for the emergent qubits, $\bra{\tau}\sigma^{[\lambda]}\ket{\tau}$ will acquire definite expectation values $\pm1$, and the coefficient $\tilde{h}_{[\mu]}$ in front of the remaining $\sigma^{[\mu]}$ term will be determined for this specific branch of the spectrum given by the Ising configuration $\ket{\tau}$:
\beq
\begin{split}
\bra{\tau}H_\text{res}\ket{\tau}&=\sum_{[\mu]}\tilde{h}_{[\mu]}\sigma^{[\mu]},\\
\tilde{h}_{[\mu]}&= \sum_{[\lambda]}h_{[\lambda][\mu]}\bra{\tau}\sigma^{[\lambda]}\ket{\tau}.
\end{split}
\eeq
In this way, a specific branching choice can be made. However, we will not make such an explicit branching choice in SBRG flow. This is in contrast to RSRG-X approach, in which one either visits each branch of the spectrum by Monte-Carlo sampling\cite{Altman:2014hg} or thermally averages over all branches\cite{Potter:2015th}. In SBRG, we keep the spectrum bifurcation structure implicitly in the Hamiltonian on the operator level, and keep track of the flow of the whole Hamiltonian in the full Hilbert space.

\begin{figure}[htbp]
\begin{center}
\includegraphics[width=220pt]{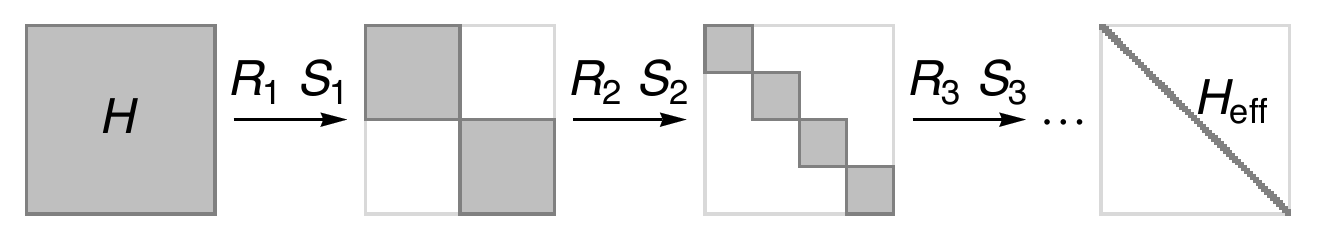}
\caption{Flow of the many-body Hamiltonian under SBRG. Each RG step consists of a Clifford group rotation $R_k$ followed by a Schrieffer-Wolff transformation $S_k$, which block diagonalizes the Hamiltonian and identifies an emergent qubit. In the end, the Hamiltonian will flow to the diagonal form $H_\text{eff}$.}
\label{fig: block diag}
\end{center}
\end{figure}

SBRG ends after the last qubit in the physical Hilbert space has been renormalized to the emergent Hilbert space. Then we are left with the effective Hamiltonian in terms of the emergent conserved quantities only
\beq\label{eq: Heff general}
H_\text{eff} = \sum_{[\lambda=0,3]}h_{[\lambda]}\sigma^{[\lambda]},
\eeq
which is fully diagonalized. So SBRG can be considered as an approximate approach to diagonalize a random many-body Hamiltonian in the strong disorder limit, as illustrated in \figref{fig: block diag}. The MBL fixed point Hamiltonian proposed in Ref.\,\onlinecite{Abanin:2013lc,Huse:2014ec,Abanin:2015io} also takes the same form as \eqnref{eq: Heff general}, and the key point there was that the emergent conserved quantities are localized in terms of the original physical qubits, which can be verified in our numerics later.

\subsection{Limitations and Strengths of SBRG}

SBRG is a generalized and improved application of RSRG-X for a class of qubit models whose spectrum bifurcates at each RG step. We should admit that SBRG is not a generic solver of disordered many-body Hamiltonians. The method has several limitations: (i)\;SBRG only works for Ising/Majorana-like qubit models which have the bifurcating spectral structure. (ii)\;Within the scope of qubit models, SBRG still suffers from the problem of Pauli-basis dependency. (iii)\;Even under a nice choice of the basis, SBRG only works in the strong disorder regime, and breaks down near the thermalization transition. Limitations (ii) and (iii) are shared by RSRG-X, where (ii) can be overcome by DMRG-X\cite{Khemani:2015qf,Yu:2015vn,Lim:2015zr,Kennes:2015yg} and other MPO-based RG\cite{Clark:wf}. We will explain these limitations in details as follows.

The current scheme of SBRG is designed to work with the qubit model \eqnref{eq: H qubit model}, and crucially relies on the assumption that there is only one non-degenerate leading energy scale in each RG step. It will not work, for example, with the models built from three-state quantum rotors, or with SU(2) symmetric spin models which have more than one terms of the same leading energy scale and singlet-triplet spectrum branching (which is not bifurcating). To run SBRG, we must assume the many-body spectrum to follow certain bifurcation structures in a $2^N$ dimensional Hilbert space, such that the RG can follow the bifurcation tree and discover the full spectrum progressively. That is why this RG scheme is called ``spectrum bifurcation''. Of course, further improvements can be made to generalize the current RG scheme to work with non-bifurcating spectrums, however we will leave this direction for future research. 

SBRG is not a basis independent RG scheme. It requires an fine-tuned Hamiltonian where the leading energy scale term at each  RG step must be given by a single Pauli operator. A local change of basis could spoil this requirement. For example if $h\sigma^{33}$ turns out to be a leading energy scale term. Under an arbitrary two-qubit basis rotation, say $e^{\ii\theta\sigma^{31}/2}$, it could be transformed to $h\cos\theta \sigma^{33}+h\sin\theta\sigma^{02}$, then each of the terms is not the true leading energy scale anymore. Thus local basis transformation could affect SBRG flow, although the physics should not be affected. So the Hamiltonian must be written in a nice basis such that the leading energy scale term is always represented as a single Pauli operator (or at least approximately). In general, local basis rotations will introduce complicated correlations among the coefficients $h_{[\mu]}$ in the qubit model \eqnref{eq: H qubit model}. So if we assume the coefficients $h_{[\mu]}$ are all independent (uncorrelated), then in the strong disorder limit, nearby Pauli operators will  have very different energy scales. In this case, the strongest Pauli operator will be a good approximation of the true leading energy scale term locally.

\begin{figure}[htbp]
\begin{center}
\includegraphics[width=140pt]{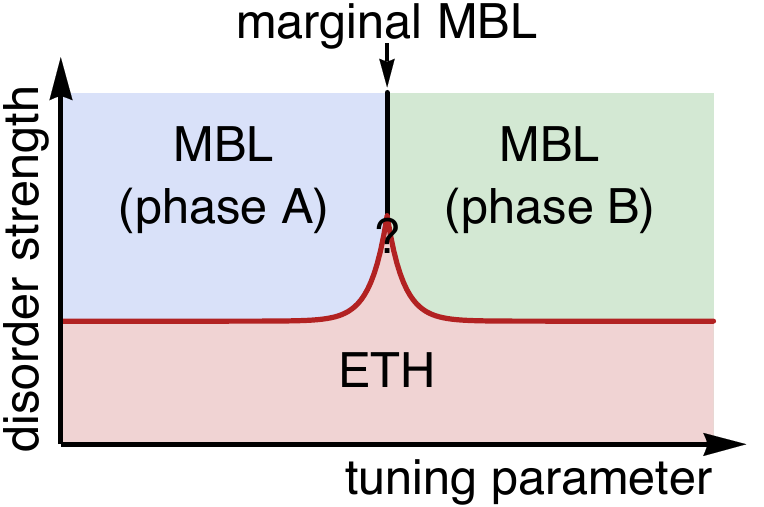}
\caption{Schematic phase diagrams containing two different kinds of transitions: the MBL-ETH transition (in red) and the MBL-MBL transition (marginal MBL, in black). At weak disorder, eigenstates in a many-body system typically thermalized. However as the disorder gets strong enough, they can be many-body localized. The MBL phase can be further divided into different quantum phases by various quantum orders. Different MBL phases are separated by the marginal MBL critical line. Whether the marginal MBL to ETH transition happens at a finite disorder strength is still an open question. SBRG fails in and near the ETH phase, but works well at the MBL-MBL transition and in the MBL phase.}
\label{fig: MBL ETH}
\end{center}
\end{figure}

As a kind of strong disorder RG, SBRG only works in the MBL phase and breaks down near the MBL-ETH transition (see \figref{fig: MBL ETH}). Two things could happen when we approach the thermalization transition from the MBL side. (i)\;Uncontrolled number of terms could be generated by the 2nd order perturbation in \eqnref{eq: H 2nd order}, such that the terms in the Hamiltonian could grow exponentially (or even faster), which would crash SBRG program. (ii)\;The off-diagonal terms $\Sigma$ become close in energy scale to the leading term $H_0$, such that the perturbative treatment is no longer valid. However if we keep away from the ETH phase, SBRG should work well in the MBL phase and on the \emph{marginal MBL} boundary\cite{Potter:2014mh} (also known as the quantum critical glass\cite{Potter:2015th}) between two MBL phases (see \figref{fig: MBL ETH}). Although delocalization happens at the marginal MBL criticality (i.e. the MBL-MBL transition\cite{Altman:2014dq,Altman:2014hg}), yet the growth rate of the Hamiltonian terms is still controlled by the MBL phases from both sides. Thus SBRG is still applicable to the marginal MBL system.

Despite of the above limitations, SBRG inherited the high efficiency of RSRG. It is fast compared to DMRG-X, and can handle relatively large system size (typical system sizes used in this work is of 128 to 512 qubits). Of course, the efficiency is gained at the cost of losing the accuracy. However if our goal is to study the overall structure of the MBL spectrum other than individual states, SBRG turns out to be an efficient method. To improve the accuracy, it is possible to pass the MPS states generated by SBRG as the initial states to DMRG-X for further improvement.

The major improvement of SBRG compared to the generic RSRG-X is to encode the spectrum branching history implicitly in the Hamiltonian flow. This allows SBRG to obtain the MBL fixed-point Hamiltonian in one run. Many physical properties of the fixed-point Hamiltonian can be therefore investigated, including the scaling and distribution of the energy coefficients and the emergent conserved quantities (to be elaborated in the following sections). SBRG can also generate an unitary MPO that approximately diagonalizes the MBL Hamiltonian. It turns out that the MPO can be approximately express as a Clifford circuit which has a high computational efficiency. With the efficient MPO, a controlled holographic mapping of the entire many-body Hilbert space becomes possible, which provides us some geometric interpretations of the entanglement structures in the MBL states. Finally as a numerical approach, SBRG allows new terms to be generated with the RG flow, which overcomes the limitation of closed-form RG (as RSRG-X was originally proposed), and enables us to study strongly interacting and higher dimensional MBL systems.

\section{Application to Quantum Ising Model}\label{sec: application}

\subsection{Beyond the Closed-Form RG}

To illustrate and to benchmark the SBRG scheme, we will take the disordered quantum Ising model (transverse field Ising model plus interaction) as an example. The model is defined on a 1D spin chain, with both XX and ZZ coupling and the external field Z,
\beq\label{eq: Ising int}
H=-\sum_{i}J_{i}\sigma_{i}^1\sigma_{i+1}^1+K_{i}\sigma_{i}^3\sigma_{i+1}^3+h_i\sigma_{i}^3.
\eeq
The model can also be interpreted as the interacting Majorana fermion chain under the Jordan-Wigner transformation,
\beq\label{eq: Majorana int}
\begin{split}
H=-\sum_{i}& \frac{J_{i}}{4}(c_{i}^\dagger c_{i+1}-c_{i}c_{i+1}+h.c.)\\
&+\frac{K_i}{4} \big(n_i-\frac{1}{2}\big) \big(n_{i+1}-\frac{1}{2}\big)- \frac{h_i}{2} \big(n_i-\frac{1}{2}\big).
\end{split}
\eeq
The coefficients $J_{i}$, $K_{i}$ and $h_{i}$ are random variables independently drawn from uncorrelated distributions. In the following, we will switch between the spin and the fermion interpretations whenever which one is more convenient. 

\begin{figure}[htbp]
\begin{center}
\includegraphics[width=240pt]{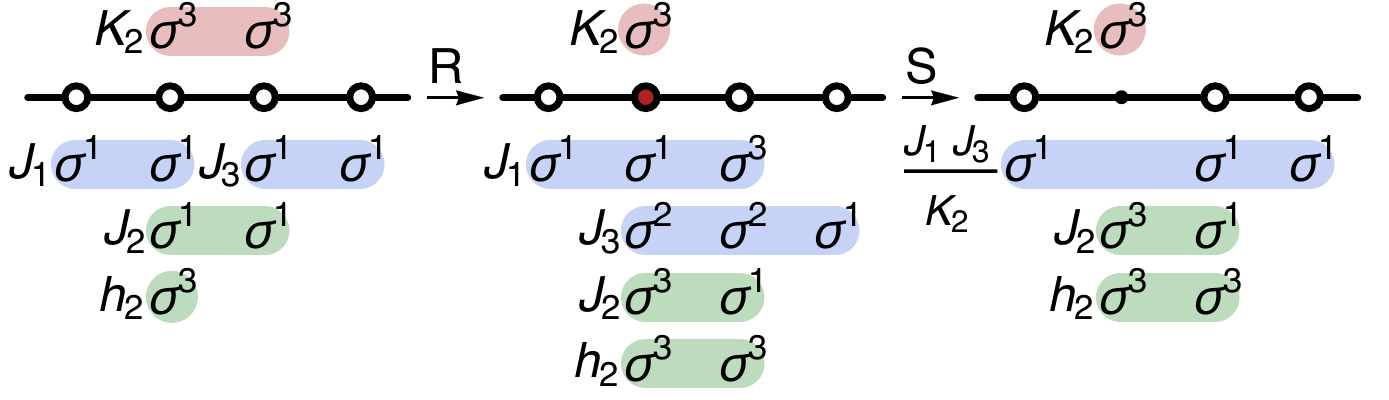}
\caption{Illustration of a single RG step, with the leading energy scale term $H_0$ in red, the block-diagonal terms $\Delta$ in green, and the block-off-diagonal terms $\Sigma$ in blue. The Hamiltonian is first transformed to the diagonal basis of $H_0$ by a Clifford rotation $R$. At this step, one qubit (marked in red) is identified as the emergent conserve quantity. Then the block-off-diagonal terms $J_1$ and $J_3$ are treated as perturbations, which generate effective couplings among the rest of the qubits via a Schrieffer-Wolff transformation $S$.}
\label{fig: RG step}
\end{center}
\end{figure}

In the original proposal\cite{Altman:2014hg} of RSRG-X, the bond term $J_i\sigma_{i}^1\sigma_{i+1}^1$ and the site term $h_i\sigma_i^3$ are considered as leading energy scales, and the ``interaction'' term $K_i\sigma_{i}^3\sigma_{i+1}^3$ is treated as perturbation. Now in SBRG, all terms are allowed to be the leading energy scale. For example, suppose $K_2$ is the leading energy scale in the following 4-site problem, one step of SBRG goes as:
\beq
\begin{split}
H&=-K_2\sigma^{0330}-J_1\sigma^{1100}-J_3\sigma^{0011}\\
&\hspace{12pt} - J_2\sigma^{0110}-h_2\sigma^{0300}-\cdots\\
&\xrightarrow{R}-K_2\sigma^{0\hl{3}00}-J_1\sigma^{1\hl{1}30}-J_3\sigma^{0\hl{2}21}\\
&\hspace{16pt} - J_2\sigma^{0\hl{3}10}-h_2\sigma^{0\hl{3}30}-\cdots\\
&\xrightarrow{S}-\Big(K_2+\frac{J_1^2+J_3^2}{2K_2}\Big)\sigma^{0\hl{3}00}-\frac{J_1J_3}{K_2}\sigma^{1\hl{0}11}\\
&\hspace{16pt} - J_2\sigma^{0\hl{3}10}-h_2\sigma^{0\hl{3}30}-\cdots,
\end{split}
\eeq
which consists of a Clifford rotation $R$ followed by a Schrieffer-Wolff transformation $S$ (as illustrated in \figref{fig: RG step}). The leading energy scale term $\sigma^{0\hl{3}00}$ is identified as a new emergent conserved quantity, and is ascribed to the effective Hamiltonian. The spectrum is also bifurcated to high- and low-energy subspaces, depending on whether the emergent qubit operator $\sigma^{\hl{3}}$ takes $+1$ or $-1$ eigenvalues. The remaining terms are passed down to the next RG step. The difference with the non-interacting case is that the residual Hamiltonian in the physical Hilbert space no longer keeps the form of \eqnref{eq: Ising int}, new terms are generated: including the interaction among more qubits like $(J_1J_3/K_2)\sigma^{1\hl{0}11}$ and the Zeeman field in other directions like $J_2\sigma^{0\hl{3}10}$. Some of the terms may contain the operator $\sigma^{\hl{3}}$ on the emergent qubit, which encodes the dependence of the spectrum branching. So in general, it is no longer possible to preserve the RG transformation in a closed form in terms of a few decimation rules. Therefore SBRG is in principle a numerical method, which can be used to explore the full MBL phase diagram of \eqnref{eq: Ising int} in the strong interaction regime.

\begin{figure}[b]
\begin{center}
\includegraphics[width=148pt]{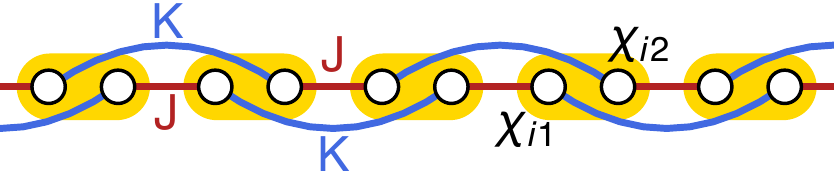}
\caption{Illustration of the model \eqnref{eq: rand XY} in terms of Majorana fermions. After the Jordan-Wigner transform, the complex fermion $c_i$ on each site (yellow block) can be split into two Majorana fermions (white circles) as $c_i=\chi_{i 1}+\ii\chi_{i 2}$. Then the model becomes two decoupled free Majorana chains.}
\label{fig: double Maj chain}
\end{center}
\end{figure}

However before running the numerics, let us mention several limits of the disordered quantum Ising model in \eqnref{eq: Ising int} which are well understood. (i) In the $K_i=0$ limit, the model simply goes back to the transverse field Ising model, and corresponds to a free Majorana chain. When $\overline{\ln J}>\overline{\ln h}$, the model is in a spin glass (SG) phase of the Ising chain (or as a topologically nontrivial phase of the Majorana chain). When $\overline{\ln J}<\overline{\ln h}$, the model is in a paramagnetic (PM) phase of the Ising chain (or as a topologically trivial phase of the Majorana chain). Both of them are MBL phases, and they are separated by a marginal MBL critical point at $\overline{\ln J}=\overline{\ln h}$, which is also known as the infinite randomness critical point. (ii) In the $h_i=0$ limit, one can make a basis rotation to rewrite \eqnref{eq: Ising int} as
\beq\label{eq: rand XY}
H=-\sum_i J_i\sigma_i^1\sigma_{i+1}^1+K_i\sigma_i^2\sigma_{i+1}^2,
\eeq 
which can then be converted to two independent copies of free Majorana chains via the Jordan-Wigner transformation as illustrated in \figref{fig: double Maj chain}. Therefore although $K_i\neq0$, the model \eqnref{eq: rand XY} is still a free fermion model, and have a marginal MBL criticality at $\overline{\ln J}=\overline{\ln K}$, which can be considered as the doubled version of the $\overline{\ln J}=\overline{\ln h}$ criticality. (iii) In the $J_i=0$ limit, the model becomes a classical Ising model with random Zeeman field,
\beq\label{eq: classical Ising}
H=-\sum_i K_i\sigma_i^3\sigma_{i+1}^3+h_i\sigma_i^3,
\eeq
which is already solved in the Ising basis. However this model can not be mapped to a free fermion model, so the interaction effect still remains (although rather trivial). The model \eqnref{eq: classical Ising} has only one phase (the PM phase), where the strong-$K$ and the strong-$h$ limits are smoothly connected without phase transition.

\subsection{Characterization of the RG Flow}

As an example, we will take the quantum Ising model in \eqnref{eq: Ising int} (or equivalently the interacting fermion model in \eqnref{eq: Majorana int}), and consider the following initial distributions of the coefficients $J_i$, $K_i$ and $h_i$,\beq\label{eq: beta dist interact} 
\begin{split}
P(J)\dd J &= \frac{1}{\Gamma_0 J}\left(\frac{J}{J_0}\right)^{1/\Gamma_0}\dd J\text{ for }J\in[0,J_0],\\
P(K)\dd K &= \frac{1}{\Gamma_0 K}\left(\frac{K}{K_0}\right)^{1/\Gamma_0}\dd K\text{ for }K\in[0,K_0],\\
P(h)\dd h &= \frac{1}{\Gamma_0 h}\left(\frac{h}{h_0}\right)^{1/\Gamma_0}\dd h\text{ for }h\in[0,h_0].
\end{split}
\eeq
The disorder strength is controlled by a single parameter $\Gamma_0$. Larger $\Gamma_0$ corresponds to stronger disorder. These initial distributions of the coefficients are expected to flow under SBRG.

The first question to ask is whether the distributions flows towards the strong disorder limit? The answer however depends on the phase of the Hamiltonian. If the Hamiltonian is in the MBL phase, SBRG will flow towards strong disorder. On the other hand, if the Hamiltonian is in the ETH phase, SBRG will flow away from the strong disorder limit.

To quantify the RG flow, we investigate the many-body Thouless parameter introduced in Ref.\,\onlinecite{Huse:2014cs,Potter:2015kl,Abanin:2015oq}. The Thouless parameter can be viewed as a many-body generalization of the Thouless conductance\cite{Thouless:1972gf} in the single-particle Anderson problem, which is the ratio of the off-diagonal resonance energy $V_{mn}$ to the diagonal level spacing, as $g=|V_{mn}|/|E_{m}-E_{n}|$, where $V_{mn}$ denotes the matrix element of a local perturbation represented in the many-body eigen basis. To give a crude estimate of the Thouless parameter in the SBRG implementation, we take out the Hamiltonian \eqnref{eq: H in RG} at each RG step: $H=H_0+\Delta+\Sigma$, which contains the leading energy scale term $H_0$ and many block-off-diagonal terms $\Sigma=\Sigma_1+\Sigma_2+\cdots$. For each block-off-diagonal term, we collect the ratio of its energy scale to the leading energy scale, and define the ratio as the Thouless parameter
\beq
g=\frac{\norm{\Sigma_i}}{\norm{H_0}}.
\eeq
This ratio lies between zero and one, and is the small parameter to control the perturbative treatment in the RG scheme. In the strong disorder limit, the leading energy scale is expected to be much larger than all the other terms in the Hamiltonian, and $g$ tends to zero. Physically this ratio also resembles the ratio of the resonance energy scale over the many-body level spacing. At each RG step, the spectrum bifurcates. The leading energy scale $\norm{H_0}$ is the amount of spectrum splitting at the current RG step, which also controls the level spacing. The block-off-diagonal terms $\Sigma_i$ flips the emergent qubit at the current RG step, which can be used to characterize the resonance energy scale.

\begin{figure}[htb]
\begin{center}
\includegraphics[width=210pt]{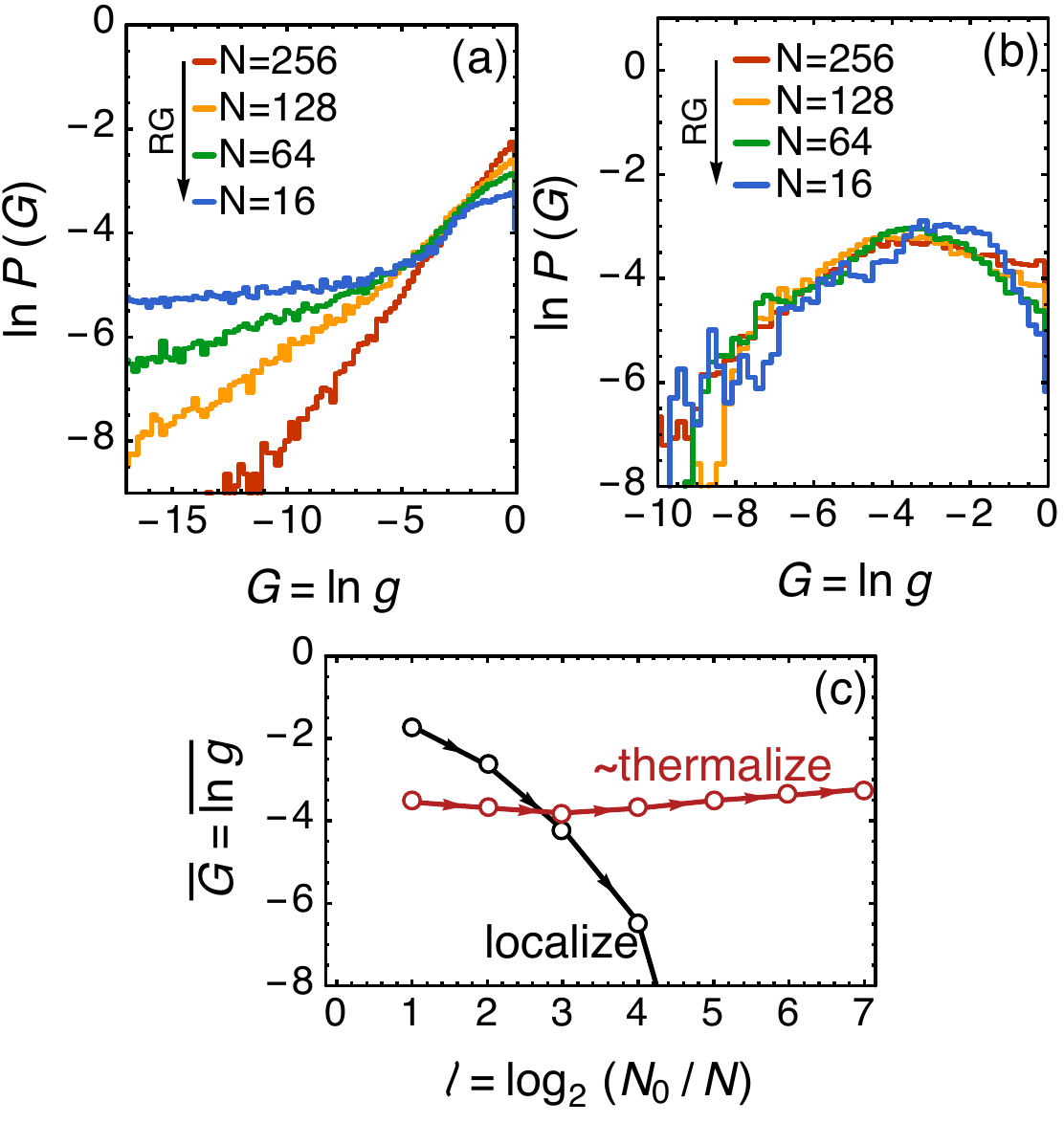}
\caption{Distribution of the logarithmic Thouless parameter $G$ (plotted in logarithmic scale) for disordered quantum Ising model at (a) $(J_0, K_0,h_0)=(1,1,1)$ and $\Gamma_0=1$, (b) $(J_0,K_0,h_0)=(2,1,1)$ and $\Gamma_0=1$. The calculation is performed on a 512-site lattice over sufficient amount of random realizations. $G$ is collected when the RG runs to the steps that the number of physical qubits is reduced to $N$. (c) Flow of the typical $\overline{G}$ under RG. $(N_0/N)$ characterizes the length scale. Smaller $N$ corresponds to longer length scale.}
\label{fig: Thouless}
\end{center}
\end{figure}

It will be more convenient to study the logarithmic Thouless parameter, denoted as $G\equiv \ln g=\ln(\norm{\Sigma_i}/\norm{H_0})$. We collect the parameter $G$ with the RG flow for various disorder realizations, from which we can study how the probability distribution $P(G)\dd G$ flows under RG. \figref{fig: Thouless}(a) shows the case that the system is in the MBL phase. The distribution $P(G)\sim e^{G/\Gamma}$ has an exponential tail in the large negative $G$ limit ($G\to-\infty$). The tail keeps broadening with the RG flow (corresponding to $\Gamma$ flowing towards $+\infty$). Therefore the average $G$ will flow to $-\infty$, or the typical value of the Thouless parameter $g$ will flow to zero, as shown in \figref{fig: Thouless}(c). This demonstrates that SBRG flows towards the strong disorder limit in the MBL phase, and become progressively accurate. In contrast, \figref{fig: Thouless}(b) shows the case that the system is close to the ETH phase (or weakly thermalized). In this case, the distribution of $G$ almost does not flow under RG, and even shifts (to the right) towards $G\to0$ near the end of the RG flow. Correspondingly average $G$ does not flow towards (or even flows away from) the strong disorder limit, as shown in \figref{fig: Thouless}(c). So the RG will eventually break down in the ETH phase, and the signature of thermalization can be indicated from the trend of the RG flow.
 
In conclusion, SBRG only works for MBL systems, and can not approach the thermalization transition or enter the ETH phase. Whether SBRG applies to the marginal MBL system is more subtle. Marginal MBL states are more delocalized than MBL states, and hence easier to thermalize if the disorder strength is weaken. It is still an open question whether marginal MBL states are stable against thermalization in the presence of interaction. However even if the marginal MBL state is unstable against thermalization, the thermalization effect is expected to be weak in the strong disorder regime. So it is still valid to talk about marginal MBL states as the short-range physics (or for finite-sized system). In this sense, SBRG is applicable to the marginal MBL system as well. Various universal scaling properties of the marginal MBL state can be studied by SBRG, as will be shown in  later sections.

\subsection{Benchmarking the Many-Body Spectrum}

At the end of the RG flow, we will obtain the full energy spectrum, encoded in the effective Hamiltonian
\beq\label{eq: Heff int}
H_\text{eff}=\sum_{i}\epsilon_i\tau_i+\sum_{i<j}\epsilon_{ij}\tau_i\tau_j+\sum_{i<j<k}\epsilon_{ijk}\tau_i\tau_j\tau_k + \cdots,
\eeq
where $\tau_i=\pm1$ is the emergent conserved quantities identified in SBRG flow. By iterating over all configurations of $\tau_i$, all the eigen energies can be obtained from \eqnref{eq: Heff int} approximately.

\begin{figure}[htbp]
\begin{center}
\includegraphics[width=125pt]{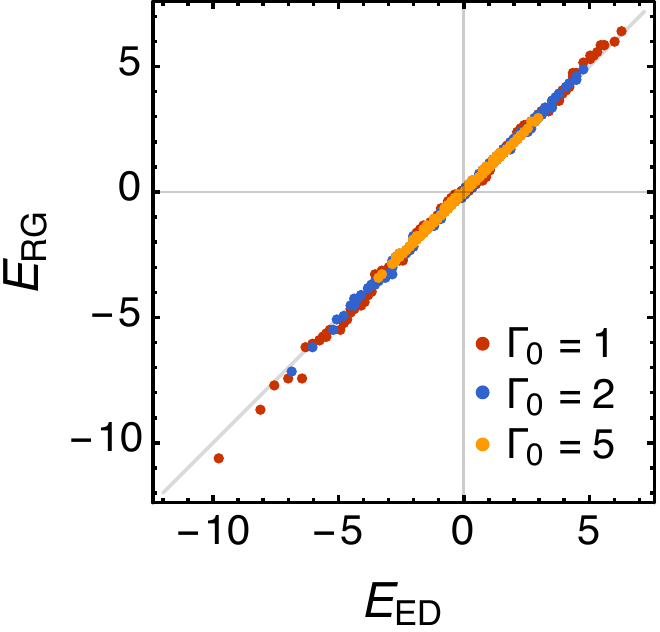}
\caption{Benchmark the many-body spectrum obtained by SBRG with ED. Calculation performed on a 8-site lattice, with $J_i$, $K_i$, $h_i$ independently drawn from the distribution in \eqnref{eq: beta dist interact} with $J_0=K_0=h_0=1$ for $\Gamma_0=1,2,5$.}
\label{fig: benchmark interact}
\end{center}
\end{figure}

The many-body energy spectrum estimated by SBRG can be benchmarked with ED on small-sized system (on which ED can be performed). In \figref{fig: benchmark interact}, we show such a comparison of the many-body spectrum in an 8-qubit system (256 eigen energies) for various shapes of the initial distributions of $h_{[\mu]}$ (controlled by $\Gamma_0=1,2,5$, where the larger $\Gamma_0$ the stronger disorder). In general, all points almost line up straightly along the diagonal line, showing good agreement between SBRG estimation and the exact spectrum from ED. Also as the disorder gets stronger (yellow points $\Gamma_0=5$), the agreement becomes better, showing that the accuracy of SBRG is systematically improved towards the strong disorder limit.

\subsection{Statistics of Energy Coefficients}

\begin{figure}[htbp]
\begin{center}
\includegraphics[width=230pt]{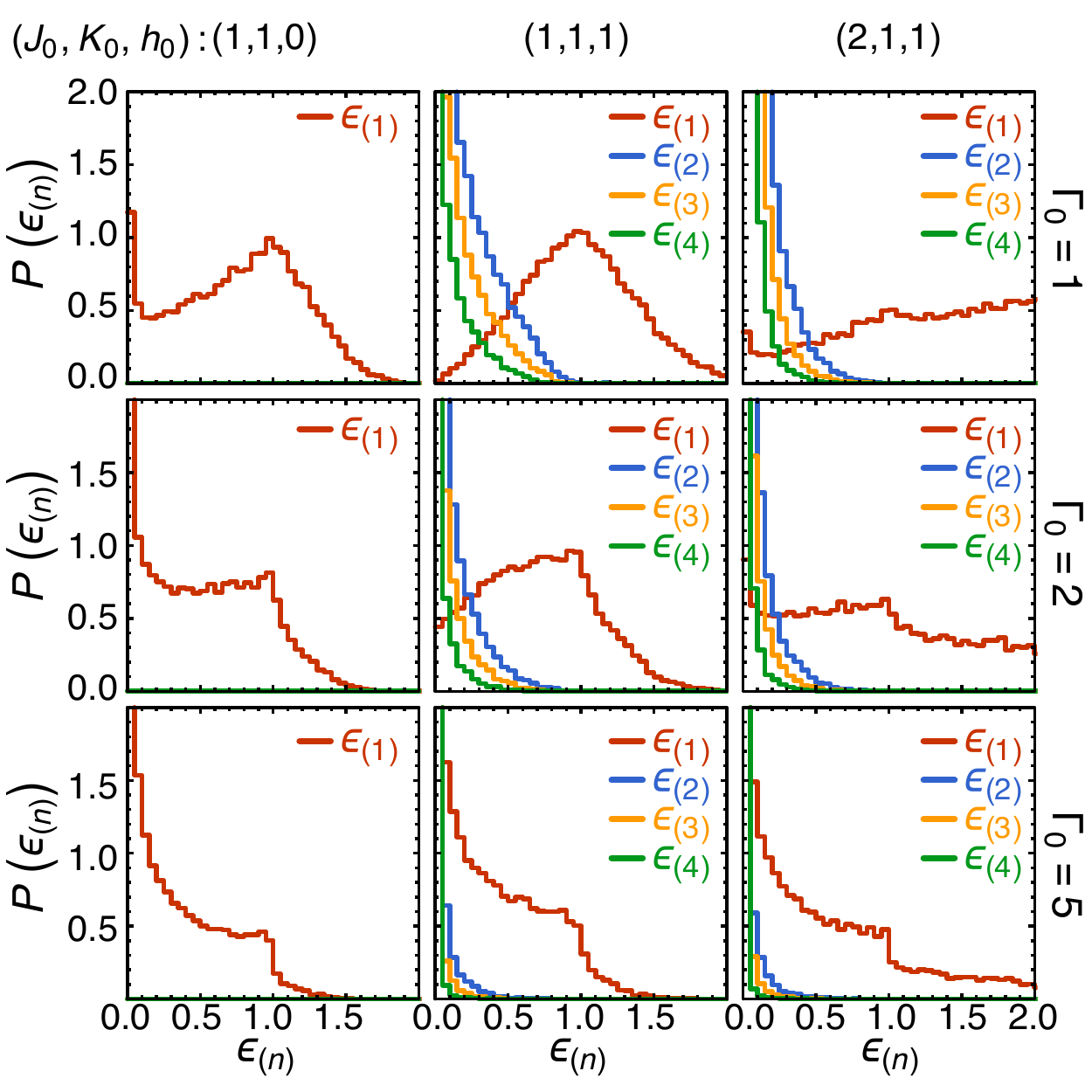}
\caption{Probability distribution of the $n$-body energy coefficients $\epsilon_{(n)}$ in the effective Hamiltonian. The statistics is collected from 200 random realizations on a 128-site lattice for each case. Different columns correspond to different settings of the initial scales $(J_0,K_0,h_0)$, and different rows correspond to different randomness of the initial distributions. For $(J_0,K_0,h_0)=(1,1,0)$, the many-body interaction terms $\epsilon_{(n)} \;(n=2,3,4,\cdots)$ are not generated, and hence their statistics are not shown. $\Gamma_0=1$ corresponds to the uniform initial distribution, and larger $\Gamma_0$ corresponds to stronger disorder.}
\label{fig: DOS interact}
\end{center}
\end{figure}

Having obtained the effective Hamiltonian in \eqnref{eq: Heff int} from SBRG, we can look at the statistics of the energy coefficients $\epsilon_i$, $\epsilon_{ij}$, $\epsilon_{ijk}$ ect. Let us denote the $n$-body coefficient $\epsilon_{i_1i_2\cdots i_n}$ as $\epsilon_{(n)}$, and study its probability distribution $P(\epsilon_{(n)})\dd\epsilon_{(n)}$. The numerical results are shown in \figref{fig: DOS interact}.

First let us look at the $(J_0,K_0,h_0)=(1,1,0)$ free fermion critical point,  mentioned in \eqnref{eq: rand XY}, as shown in the left column of \figref{fig: DOS interact}. In this case, the many-body terms $\epsilon_{(n)}\;  (n=2,3,4,\cdots)$ are not generated under the RG flow, showing that SBRG has automatically found the single particle basis to diagonalize the Hamiltonian. The distributions of the single particle levels $\epsilon_{(1)}$ exhibit the Dyson singularity\cite{Dyson:1953pd,Rieder:2014ty} at low energy (or low frequency), i.e.\,$P(\epsilon_{(1)})\sim \epsilon_{(1)}^{-1}(\ln(W/\epsilon_{(1)}))^{-3}$ where $W$ is the typical band width.

The middle column of \figref{fig: DOS interact} shows the cases of $(J_0,K_0,h_0)=(1,1,1)$, which is in the PM (or SPT trivial) MBL phase according to phase diagram \figref{fig: phase diagram} (to be discussed later). For the weak disorder case $\Gamma_0=1$, the single-particle DOS (the distribution of $\epsilon_{(1)}$) drops to zero at the low frequency limit ($\epsilon_{(1)}\to 0$), which is in reminiscence of the fermion single-particle band gap in the clean limit. As the disorder gets stronger ($\Gamma_0=2,5$), the band gap is gradually filled up by the in-gap localized states. The right column of \figref{fig: DOS interact} shows the cases of $(J_0,K_0,h_0)=(2,1,1)$, which sits right at the interacting marginal MBL critical line. The single-particle DOS remains gapless in these cases, which is consistent with the fact that the single-particle gap must close at the transition between the topological ($\nu=1$) and the trivial ($\nu=0$) 1D fermion SPT phases,\cite{Kitaev:2009sd,Fidkowski:2010iv,Fidkowski:2011pa} even in the presence of disorder and interaction.

Now let us turn to the many-body terms $\epsilon_{(n)}\; (n=2,3,4\cdots)$. Away from the free fermion limit, the many-body terms will appear under the RG flow, and the effective Hamiltonian $H_\text{eff}$ becomes an interacting one, as shown in the middle and right columns of \figref{fig: DOS interact}. In the strong disorder limit $\Gamma_0=5$, the distributions look similar to the free fermion case, and little interaction is generated among the emergent conserved quantities. As the disorder gets weaker, the interaction becomes stronger (as the distribution shifts towards higher energy scale) and involves more emergent conserved quantities. Many of the interaction terms are UV-IR mixing terms, which describe how the spectrum branching in the (high-frequency) UV  limit could shift and rearrange the energy levels in (low-frequency) IR limit.

\begin{figure}[htbp]
\begin{center}
\includegraphics[width=114pt]{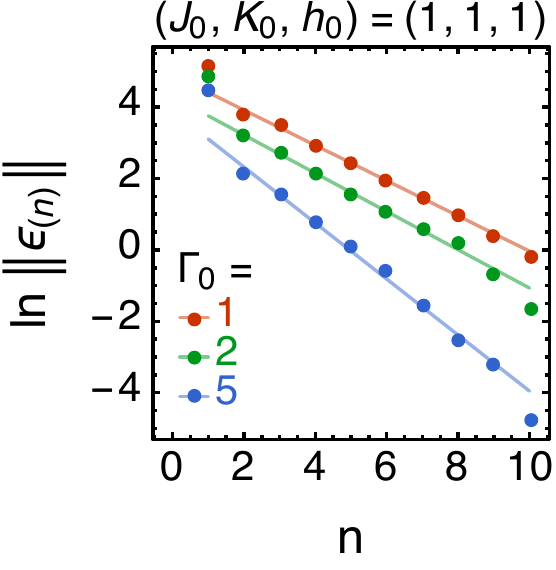}
\caption{The energy scale $\norm{\epsilon_{(n)}}$ of $n$-body terms decays with $n$ exponentially in the MBL phase. The decay rate depends on the initial randomness $\Gamma_0$.}
\label{fig: n scaling}
\end{center}
\end{figure}

The many-body energy coefficients $\epsilon_{(n)}$ also set the energy scale of many-body resonances in the MBL system. To quantify the $n$-body resonance energy scale, we studied the norm of the $n$-body terms $\norm{\epsilon_{(n)}}$, defined as
\beq
\norm{\epsilon_{(n)}}=\sqrt{\int \epsilon_{(n)}^2P(\epsilon_{(n)})\dd\epsilon_{(n)}}.
\eeq
We found that the $n$-body resonance energy scale decays with $n$ exponentially,
\beq
\norm{\epsilon_{(n)}}\sim We^{-n/\zeta},
\eeq
as shown in \figref{fig: n scaling}. This observation justifies the same proposal in Ref.\,\onlinecite{Huse:2015pw}. From the exponential decay of the $n$-body resonance energy scale, it was further argued\cite{Huse:2015pw} that the conductivity should follow a power-law of frequency $\sigma(\omega)\sim\omega^{2-\phi}$ in the deep MBL phase. From \figref{fig: n scaling}, one can also observe that as the disorder gets weaker (smaller $\Gamma_0$), $\norm{\epsilon_{(n)}}$ decays slower with $n$, meaning that the many-body interaction/resonance will become more important for weaker disorder.

In the strong disorder and weak interaction limit, where the many-body terms are suppressed, the effective Hamiltonian $H_\text{eff}=\sum_{i}\epsilon_{i}\tau_{i}$ is expected to produce the Poisson level statistics due to the lack of level repulsion, which is consistent with the MBL physics. Away from that limit, the many-body terms can reshuffle the energy levels at low-frequency, which may change the level statistics drastically. However, it is not clear to us if the level statistics imposes any constraint on the distributions $P(\epsilon_{(n)})$ of the energy coefficients, or if it is possible to determine the level statistics from $P(\epsilon_{(n)})$. The discussion along this line may go beyond the scope of this work, and we will leave these interesting questions for future study.

\subsection{RG Transform as Clifford Circuit}

SBRG is a Hilbert space preserving RG. The RG transformation contains no isometry, but only unitary transforms. So by collecting the unitary transforms that have been performed in each RG step, we can obtain a unitary mapping which maps the physical Hilbert space to the emergent Hilbert space, and consequently maps the initial Hamiltonian $H$ to the effective Hamiltonian $H_\text{eff}$ (approximately). In each RG step, the Hamiltonian is conjugated by two unitary transforms: a Clifford group rotation $R_k$ followed by a Schrieffer-Wolff transformation $S_k$ (here $k=1,2,\cdots$ labels the $k$th RG step). Therefore the whole RG transformation is a product of them as 
\beq\label{eq: U as RS} U_\text{RG}=R_1S_1R_2S_2\cdots=\prod_{k}R_kS_k,\eeq
which approximately diagonalize the many-body Hamiltonian $H\to H_\text{eff}\simeq U_\text{RG}^\dagger H U_\text{RG}$. Note that in SBRG algorithm, the Schrieffer-Wolff transformation $S_k$ is not performed exactly, but only carried out to the 2nd order in the perturbative expansion, where all the higher order terms are truncated. So the truncation error can build up in the Hamiltonian under the RG flow, and hence the resulting RG transform $U_\text{RG}$ is only an approximate unitary transform to diagonalize the Hamiltonian. Nevertheless the approximation is expected to be good if the disorder is strong (such as in the full MBL phase).

Therefore the RG transform $U_\text{RG}$ actually encodes all the many-body eigenstates of $H$ to some approximation. To retrieve the eigenstates from $U_\text{RG}$, we just need to take the eigenstates of the effective Hamiltonian $H_\text{eff}$, which are direct product states $\ket{\{\tau_i\}}=\ket{\tau_1}\otimes\ket{\tau_2}\otimes\cdots$ of the emergent qubits $\tau_i=\pm1$, and transform them back to the original basis by reversing the RG transformation
$\ket{\Psi_{\{\tau_i\}}}\simeq U_\text{RG}\ket{\{\tau_i\}}$. In this sense, $U_\text{RG}$ can be viewed as a quantum circuit\cite{Nielsen:2000qr} which prepares the entangled eigenstate from the disentangled product state. Different energy eigenstates $\ket{\Psi_{\{\tau_i\}}}$ in the MBL spectrum correspond to different input product state $\ket{\{\tau_i\}}$ of the emergent qubits.

\begin{figure}[htbp]
\begin{center}
\includegraphics[width=216pt]{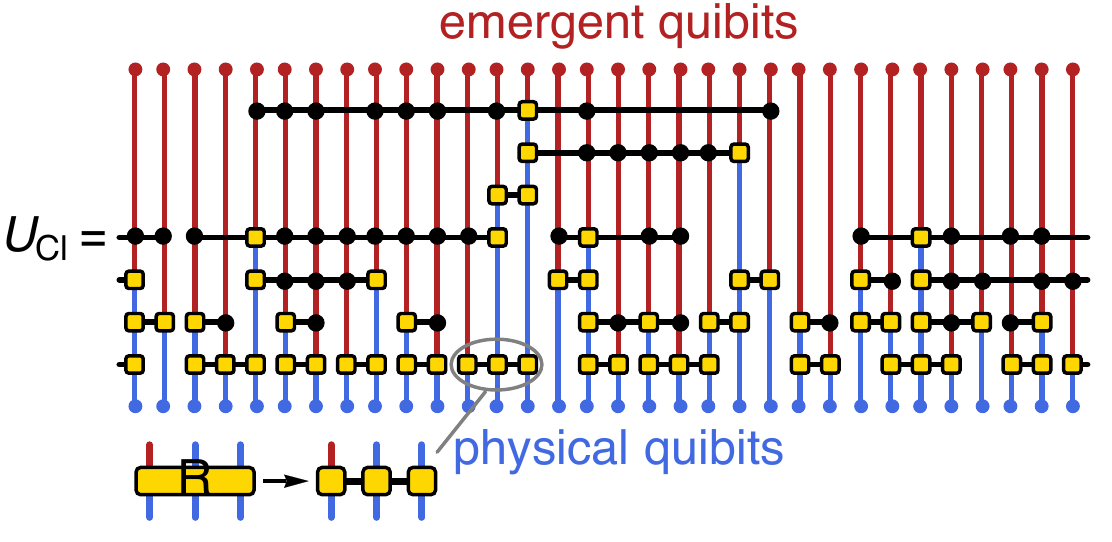}
\caption{The random Clifford circuit $U_\text{Cl}$. Each Clifford rotation $R_k$ is viewed as a (controlled) quantum gate, where the black dots mark the control qubits and the yellow squares are generic unitary gates. Under the action of each gate, one physical qubit (in blue) will be transformed to one emergent qubit (in red). Only the emergent qubit serves as the control.}
\label{fig: circuit}
\end{center}
\end{figure}

Although we have the Clifford rotation $R_k$ and SW transformation $S_k$ matrices in hand, computing their product explicitly as in \eqnref{eq: U as RS} is still  expensive in numerics. So at this stage, we  make further approximations by dropping all the SW transformations $S_k$, and reconstruct the eigenstates just from the Clifford rotations $R_k$, because $S_k$ are expected to be close to identity. With $R_k$ only, the remaining unitary transformation is a quantum circuit of random Clifford gates, dubbed as the random Clifford circuit, or the random stabilizer circuit\cite{Gottesman:1998cl,Gottesman:1998lk,Chandran:2015rc}
\beq U_\text{Cl}=\prod_k R_k.\eeq
Each Clifford gate $R_k$ here is a controlled gate in general, which transforms one physical qubit to one emergent qubit under the control of the previously identified emergent qubits, as illustrated in \figref{fig: circuit}. The emergent qubits will not be further rotated by the later gates, but may serve as the control qubits for the later gates. The Clifford circuit is a further approximation of the RG transformation, which gives cruder estimate of the eigenstate
\beq\label{eq: RC state} \ket{\Psi_{\{\tau_i\}}}\simeq U_\text{Cl}\ket{\{\tau_i\}}.\eeq
The accuracy is traded for efficiency. Since $U_\text{Cl}\ket{\{\tau_i\}}$ is a stabilizer state, many physical properties can be efficiently calculated using the stabilizer formalism.

It worth mention that dropping the Schrieffer-Wolff transformations $S_k$ only affects the accuracy of the estimated eigenstates, but not the SBRG flow, because the RG flow is purely guided by the flow of the Hamiltonian without referring to $S_k$. In particular, the 2nd order perturbation can be performed on the Hamiltonian level according to \eqnref{eq: H 2nd order} without explicitly calculating the Schrieffer-Wolff transformation $S$ in \eqnref{eq: SW transform}. So we can first determine the Hamiltonian flow under SBRG, and then reconstruct the Clifford circuit $U_\text{Cl}$ independently after the RG flow.

The Clifford circuit actually provides the MPS representation for all eigenstates. Because each Clifford gate can be represented as a matrix product operator (MPO)\cite{Verstraete:2004xj} with fixed bond dimension 2. Then the entire Clifford circuit is a large MPO with total bound dimension bounded by the circuit depth. As a RG transformation, the circuit depth is typically at most logarithmic in system size. The logarithmic depth is only saturated for the marginal MBL system. For systems deep in the MBL phase, the RG will terminate at finite depth, and the resulting MPO actually has a (smaller) bound dimension independent of the system size.\cite{Clark:2014ed} Inputting a direct product state from the emergent qubit side corresponds to fixing the emergent legs of the MPO (red legs in \figref{fig: circuit}), which turns the MPO to an MPS representation of the MBL eigenstate. Thus once the random Clifford circuit is generated from the SBRG flow, we have also obtained the (approximate) MPS representation for every energy eigenstate in the whole spectrum. Because the bipartite entanglement entropy is bounded by the bound dimension, the entanglement entropy can scale logarithmically at marginal MBL, and follows area-law in the MBL phase.\cite{Bauer:2013al,Swingle:2013oy}

\subsection{Benchmarking the Many-Body States}

\begin{figure}[htbp]
\begin{center}
\includegraphics[width=160pt]{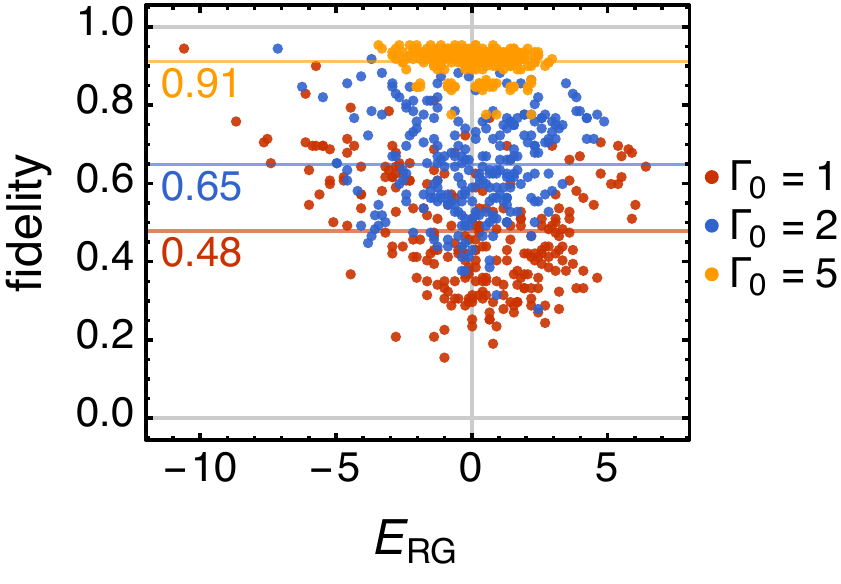}
\caption{Many-body fidelity of the energy eigenstates obtained by the Clifford circuit benchmarked with ED. Calculation performed on a 8-site lattice, with $J_i$, $K_i$, $h_i$ independently drawn from the distribution in \eqnref{eq: beta dist interact} with $J_0=K_0=h_0=1$ for $\Gamma_0=1,2,5$. }
\label{fig: fidelity}
\end{center}
\end{figure}

To check how good is the approximation of the random Clifford circuit, we can benchmark the circuit with the exact diagonalization (ED) unitary transformation. We take the quantum Ising model in \eqnref{eq: Ising int} on a 8-site lattice. First we run SBRG and collect the Clifford rotations $R_k$ ($k=1,\cdots,8$). Then we assemble the Clifford circuit $U_\text{Cl}=\prod_{k=1}^{8} R_k$ and retrieve the eigenstates from \eqnref{eq: RC state}. As we enumerate over all configurations of the emergent conserved quantities $\{\tau_i\}$, we can obtain all the approximate eigenstates $\ket{\Psi_{\{\tau_i\}}}$. Finally we overlap these states with the exact energy eigenstates obtained by ED. Typically one can order the eigenstates by their eigen energies. But since the energy spectrum estimated by SBRG also contains small error, a few eigenstates might be reshuffled. Therefore we match the corresponding eigenstates by the maximal overlap (implemented by the maximal bipartite matching algorithm\cite{Fredman:1987te}), and calculate the fidelity $f_{\{\tau_i\}}=|\langle\Psi_n^\text{ED}|\Psi_{\{\tau_i\}}\rangle|$, where $\ket{\Psi_n^\text{ED}}$ is the ED eigenstate that matches. The fidelity of each RG-generated eigenstate is plotted against its energy in \figref{fig: fidelity}. For moderate randomness $\Gamma_0=1$, the fidelity is typically around 0.5, and improves systematically as the randomness gets stronger (see the $\Gamma_0=5$ data). It is fair to say the result is acceptable for the wave function overlap in the 256-dimensional many-body Hilbert space.  The fidelity level of SBRG is also comparable to that of RSRG-X (as benchmarked in Ref.\,\onlinecite{Altman:2014hg}), showing that SBRG has a similar performance as RSRG-X. Admittedly, SBRG is a fast but less accurate method. Other approaches such as DMRG-X\cite{Khemani:2015qf,Yu:2015vn,Lim:2015zr,Kennes:2015yg} or tensor network optimization\cite{Clark:2014ed,Pollmann:2015vn} can achieve much higher eigenstate fidelity. If high-quality eigenstates are desired, the eigenstate estimated by SBRG can be passed as the initial MPS state to DMRG-X for further refinement. We also observe from \figref{fig: fidelity} that the fidelity is typically higher for the ground state and top state in the spectrum, and lower for the states in the middle of the spectrum. This is because the middle state has a stronger tendency to thermalize, and we know that SBRG and the Clifford circuit will become inaccurate towards thermalization. In conclusion, as long as we are in the MBL (or marginal MBL) phase, the Clifford circuit can provide reasonable approximation for the many-body eigenstates in the whole spectrum.

\begin{figure}[htbp]
\begin{center}
\includegraphics[width=140pt]{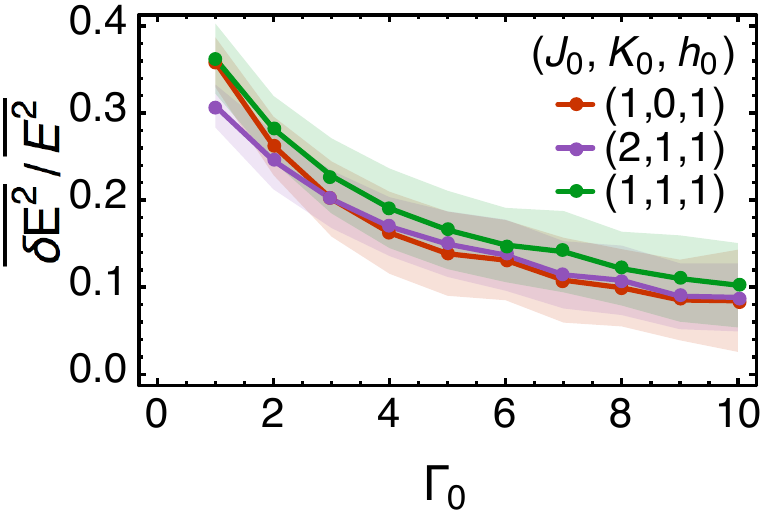}
\caption{Normalized energy variance $\overline{\delta E^2}/\overline{E^2}$ v.s. the randomness $\Gamma_0$ of the initial distribution, for various parameters $(J_0,K_0,h_0)$. The calculation is performed on a 64-site lattice with 100 random realizations for each data point.}
\label{fig: E variance}
\end{center}
\end{figure}

For larger-sized system where the full-spectrum ED is no longer available, we can check the accuracy of the Clifford circuit by the energy variance, as proposed in Ref.\,\onlinecite{Pollmann:2015vn}. We define the energy variance averaged over the approximate eigenstates as
\beq\label{eq: delta E2}
\overline{\delta E^2} =\frac{1}{2^N}\sum_{\{\tau_i\}}\bra{\Psi_{\{\tau_i\}}}H^2\ket{\Psi_{\{\tau_i\}}} - \bra{\Psi_{\{\tau_i\}}}H\ket{\Psi_{\{\tau_i\}}}^2,
\eeq
where $H$ is the initial model Hamiltonian. If $\ket{\Psi_{\{\tau_i\}}}$ were exact eigenstates, the mean energy variance $\overline{\delta E^2}$ should vanish. So the non-vanishing $\overline{\delta E^2}$ indicates how far the Clifford circuit deviates from the exact diagonalization. However, $\overline{\delta E^2}$ grows with the system size because it has the dimension of the total energy squared and the total energy is an extensive quantity that scales with the system size. To get rid of the system size dependence, we choose to normalize $\overline{\delta E^2}$ by the total variance of the energy
\beq\label{eq: E2}
\overline{E^2}=\frac{1}{2^{N}}\sum_{\{\tau_i\}}\bra{\Psi_{\{\tau_i\}}}H^2\ket{\Psi_{\{\tau_i\}}},
\eeq
and the result is shown in \figref{fig: E variance}. For various model parameters $(J_0,K_0,h_0)$, the normalized mean energy variance $\overline{\delta E^2}/\overline{E^2}$ is controlled below $\sim 0.4$,\footnote{For comparison, $\overline{\delta E^2}/\overline{E^2}$ on random states will be close to 1.} and decreases all the way to zero towards the strong disorder limit where the Clifford circuit becomes exact.

In conclusion, SBRG generates two important sets of data:  a fixed-point Hamiltonian $H_\text{eff}$ which gives the full energy spectrum to the 2nd order perturbation, and a Clifford circuit $U_\text{Cl}$ which encodes all the eigenstates to the 0th order perturbation. Admittedly, the Clifford circuit may be a crude approximation in terms of encoding the many-body eigenstates, but it has nice properties to enable highly efficient calculation of many physical quantities, which will be discussed in the following section.

\section{Entanglement Holographic Mapping}\label{sec: EHM}

\subsection{SBRG and Holography}

\begin{figure}[htbp]
\begin{center}
\includegraphics[width=220pt]{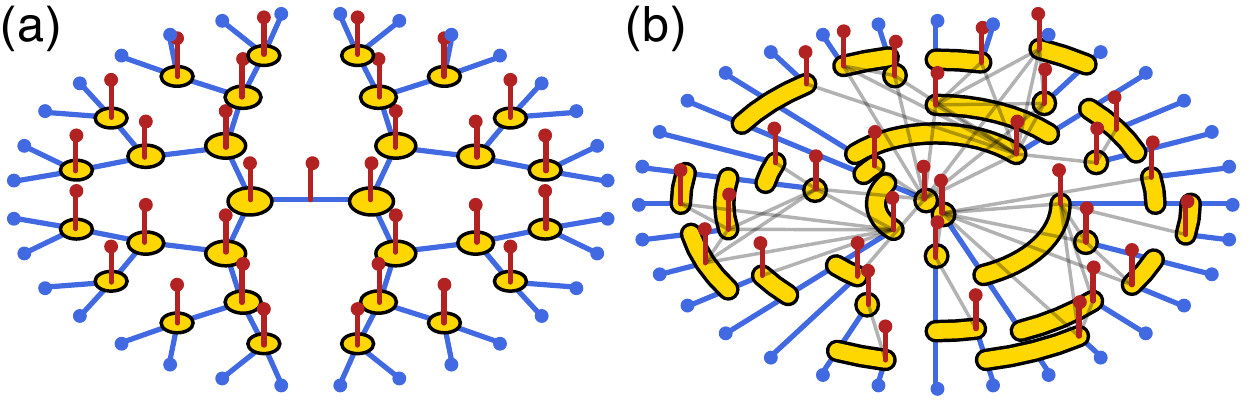}
\caption{EHM network of (a) free fermion system without disorder,\cite{Qi:2013fm} and (b) MBL system. Both EHM networks map the physical qubits (in blue) on the holographic boundary to the emergent qubits (in red) in the holographic bulk. The unitary transform in each RG step is represented by a yellow block. Each RG step will identify a new bulk qubit.}
\label{fig: EHM}
\end{center}
\end{figure}

SBRG also provides a holographic interpretation of the MBL system. This follows from the idea that every Hilbert-space-preserving RG can be interpreted as a holographic mapping, and the RG transformation is a unitary mapping from the holographic boundary to holographic bulk. The bulk qubits defined by this mapping can be considered as local degrees of freedom in an emergent bulk geometry, whose radial direction (or the holographic extra dimension) corresponds to the energy (frequency) scale, running from UV to IR, as illustrated in \figref{fig: EHM}. The close relation between RG and holography has been actively studied in both condensed matter physics and high energy physics communities.\cite{Swingle:2012yq,Swingle:2012bs,Takayanagi:2012vm,Balasubramanian:2013xz,Verstraete:2013do,Maldacena:2013te,Lee:2014qh,Takayanagi:2014an,Leigh:2014rb,Vidal:2008zp,Vidal:2014oz,Lee:2015aa,Molina-Vilaplana:2015lr,Preskill:2015ja,Takayanagi:2015so,Takayanagi:2015pd,Czech:2015ig,Bao:2015hf} SBRG is also a Hilbert-space-preserving RG, so it must also have a holographic interpretation.

Here we will follow the construction of EHM proposed by one of the authors in Ref.\,\onlinecite{Qi:2013fm,Qi:2015ct}. The construction was first applied to a free fermion system without disorder. The model describes spinless fermions hopping on a 1D lattice. At each RG step, two neighboring sites are coarse grained to an effective site either in the low-energy subspace or in the high-energy subspace, depending on the state of the bulk fermion (either empty or occupied). The coarse graining is implemented as unitary transformations of the fermion basis, represented as small yellow disks in \figref{fig: EHM}(a). In this way, a tensor network is generated under RG, which maps the boundary fermions to the bulk.

Now for SBRG, RG steps are implemented by Clifford gates (approximately), depicted as yellow blocks in \figref{fig: EHM}(b). Each Clifford gate identifies an emergent qubit on one of its output leg. We pull out the emergent qubits as bulk degrees of freedom, because they are the ones who control the spectrum branching at each RG step.  The bulk qubits reside at the radius position determined by their energy scales at which they are identified by SBRG. The energy scale can be easily read off from the effective Hamiltonian $H_\text{eff} = \sum_{i}\epsilon_i\tau_i+\cdots$, such that the energy scale associated to $\tau_i$ is simply $\epsilon_i$, i.e. the single-body energy coefficients in $H_\text{eff}$. In this way, the random Clifford circuit is interpreted as the EHM network. Since we are dealing with disordered systems, our EHM network is not a regular tensor network, and the bulk geometry also inherits the randomness on the boundary. The tensor network maps the MBL eigenstates on the boundary to a direct product state in the bulk, so it is also a disentangler network or a random version of MERA,\cite{Vidal:2008zp} which progressively removes the entanglements in the wave function.

An important feature of the SBRG-generated EHM network is that the transformations performed in the IR region are controlled by the states of the emergent qubits in the UV region, as depicted in \figref{fig: EHM}(b) by the gray lines connecting the IR gates to the UV emergent qubits. The emergent qubit state in the UV region determines the large-scale spectrum branching and roughly locates the system around certain energy density in the spectrum, therefore the IR transformations can be affected by the choice of the qubit state that has been made in the UV layer. Correspondingly, the spectrum branching dependence of the IR physics is reflected in the many-body terms $\sum_{ij}\epsilon_{ij}\tau_i\tau_j+\sum_{ijk}\epsilon_{ijk}\tau_i\tau_j\tau_k+\cdots$ in the effective Hamiltonian as a UV-IR mixing effect.

The idea of holography is useful, because it provides a geometric interpretation of the entanglement structures in the quantum many-body state. For example, the entanglement entropy can be interpreted as the minimal surface in the holographic bulk following the Ryu-Takayanagi formula.\cite{Ryu:2006fj} The correlation function or mutual information is also related to the holographic geodesic distance. In general, a full-spectrum holographic mapping for generic many-body system is challenging. However the MBL systems are special, in the sense that they are ``quasi-solvable'', which allows a Hilbert-space-preserving RG and a controlled holographic mapping of the entire many-body Hilbert space. In the following we will discuss several physical properties of MBL systems, and the holographic geometry in the bulk. 

\subsection{Stablizer Properties}

\begin{figure}[htbp]
\begin{center}
\includegraphics[width=175pt]{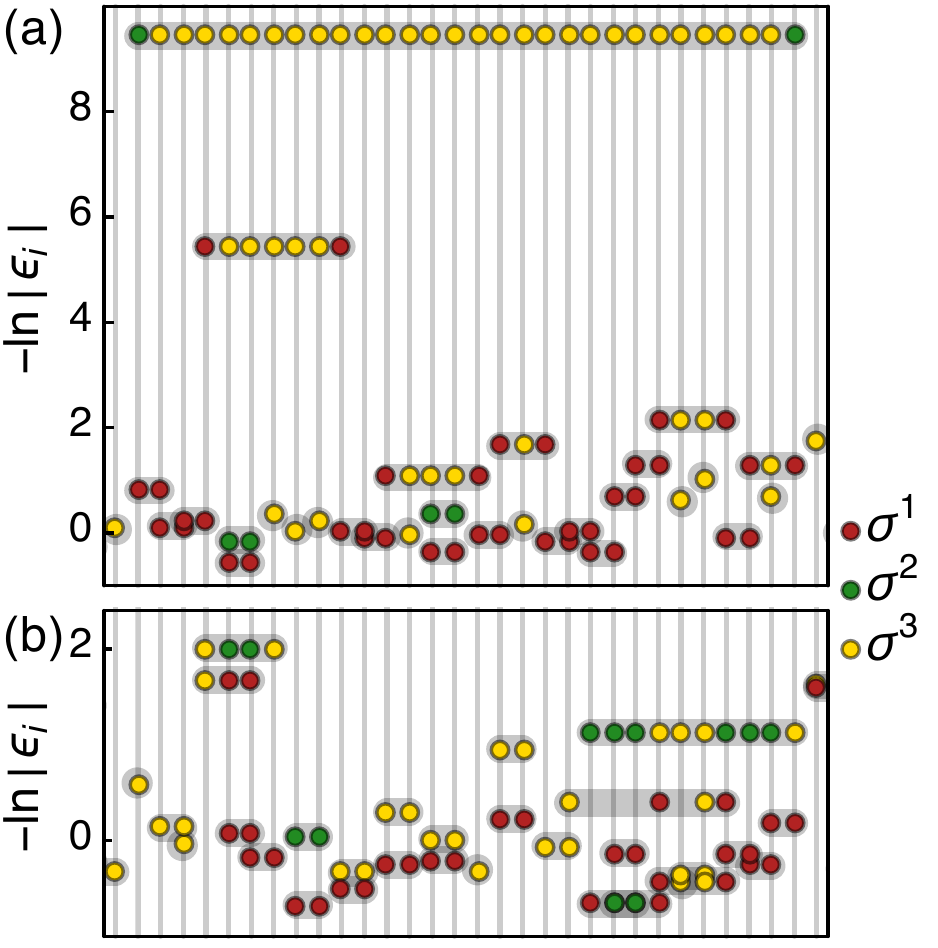}
\caption{Snapshots of the stabilizers $\hat{\tau}_i$ arranged vertically by their energy scales $\epsilon_i$ from UV (bottom) to IR (top), for (a) $(J_0,K_0,h_0)=(1,0,1)$ and (b) $(J_0,K_0,h_0)=(1,1,1)$. Horizontal axis is the real-space position in a 32-site subsystem taken from the 256-site lattice. Each stabilizer is represented as a product of Pauli matrices (grouped in gray shadows), and each Pauli matrix is illustrated by a color dot (red$=\sigma^1$, green$=\sigma^2$, yellow$=\sigma^3$).}
\label{fig: stabilizers}
\end{center}
\end{figure}

Let us first look at the emergent conserved quantities $\tau_i$, also known as the localized bits\cite{Huse:2014ec} or the local integrals of motion\cite{Abanin:2013lc,Pollmann:2015vn}, which play important roles in the phenomenology of MBL systems. On one hand, they label the emergent qubit states in the holographic bulk. On the other hand, they are the \emph{stabilizers} (commuting projectors) for the eigenstates on the holographic boundary. The representation of the emergent conserved quantities in the physical Hilbert space $\hat{\tau}_i$ can be approximately found by inversely applying the Clifford circuit,
\beq\label{eq: stabilizer by RC}
\hat{\tau}_i=U_\text{Cl}\,\sigma^{[3_i]}\,U_\text{Cl}^\dagger,
\eeq
where $[3_i]$ denotes the set of Pauli indices of the form $[0\cdots]3[0\cdots]$ with 3 appears at the $i$th qubit position for $i=1,2,\cdots,N$. We use a hat to emphasize that the operator $\hat{\tau}_i$ is represented in the physical Hilbert space, acting on the holographic boundary. From the effective Hamiltonian $H_\text{eff}$ in \eqnref{eq: Heff int}, we know that every eigenstate in the physical Hilbert space is a stabilizer state (approximately), stabilized by $\hat{\tau}_i$ to the value of the emergent conserved quantity $\tau_i=\pm 1$
\beq\label{eq: stabilizer state}
\forall i:\hat{\tau}_i\ket{\Psi_{\{\tau_i\}}}=\tau_i\ket{\Psi_{\{\tau_i\}}}.
\eeq
Each stabilizer $\hat{\tau}_i$ is also a Pauli operator, because the Clifford circuit can only take the Pauli operator $\sigma^{[\lambda]_i}$ to another Pauli operator. To gain some intuition of the stabilizers, let us plot them in \figref{fig: stabilizers}. We start with the quantum Ising model in \eqnref{eq: Ising int} (or as the fermion model in \eqnref{eq: Majorana int}), and take the random distribution from \eqnref{eq: beta dist interact} with the parameters $(J_0,K_0,h_0)=(1,0,1)$ for \figref{fig: stabilizers}(a) and $(J_0,K_0,h_0)=(1,1,1)$ for \figref{fig: stabilizers}(b). The stabilizers are collected along SBRG flow, and restored to the physical Hilbert space by \eqnref{eq: stabilizer by RC}.

One important question is the locality of these stabilizers. As we can see from \figref{fig: stabilizers}, most of them are quite well localized. \figref{fig: stabilizers}(a) shows the scenario at the marginal MBL criticality (at the critical point of free Majorana/Ising chain). In the UV regime (i.e. $-\ln|\epsilon_i|\sim 0$), many stabilizers are of the form $\sigma^{[0\cdots]11[0\cdots]}$ (two red dots) or $\sigma^{[0\cdots]3[0\cdots]}$ (a single yellow dot), corresponding to the strong bond or the strong site operators respectively. As the RG flows towards the IR limit (larger $-\ln|\epsilon_i|$, lower frequency), longer stabilizers are found, corresponding to the longer-range coupling of the Majorana fermions. The fermion Jordan-Wigner string can also be observed as $\sigma^{[3...]}$ (a line of yellow dots between the red/green dots). These long-range couplings are quite rare and have very low energy scale, which follow the universal scaling behavior $-\ln|\epsilon_i|\sim \sqrt{l_i}$ with $l_i$ being the length of the stabilizer.\cite{Fisher:1995cr} We can tune the system away from the criticality to the MBL phase by the fermion interaction $K_i(n_i-\frac{1}{2})(n_{i+1}-\frac{1}{2})$. Because on the mean-field level, the interaction effectively enhances the randomness of the site field as $h_i\to h_i+K_i(n_{i+1}-\frac{1}{2})$, and therefore tips the balance between the bond $J_i$ and the site $h_i$ terms.\cite{Altman:2014hg} As shown in \figref{fig: stabilizers}(b), with $(J_0,K_0,h_0)=(1,1,1)$, all the stabilizers become short-range and concentrated near the UV regime. We see that SBRG stops flowing within a few orders of the energy scale in the MBL phase.

\begin{figure}[htbp]
\begin{center}
\includegraphics[width=200pt]{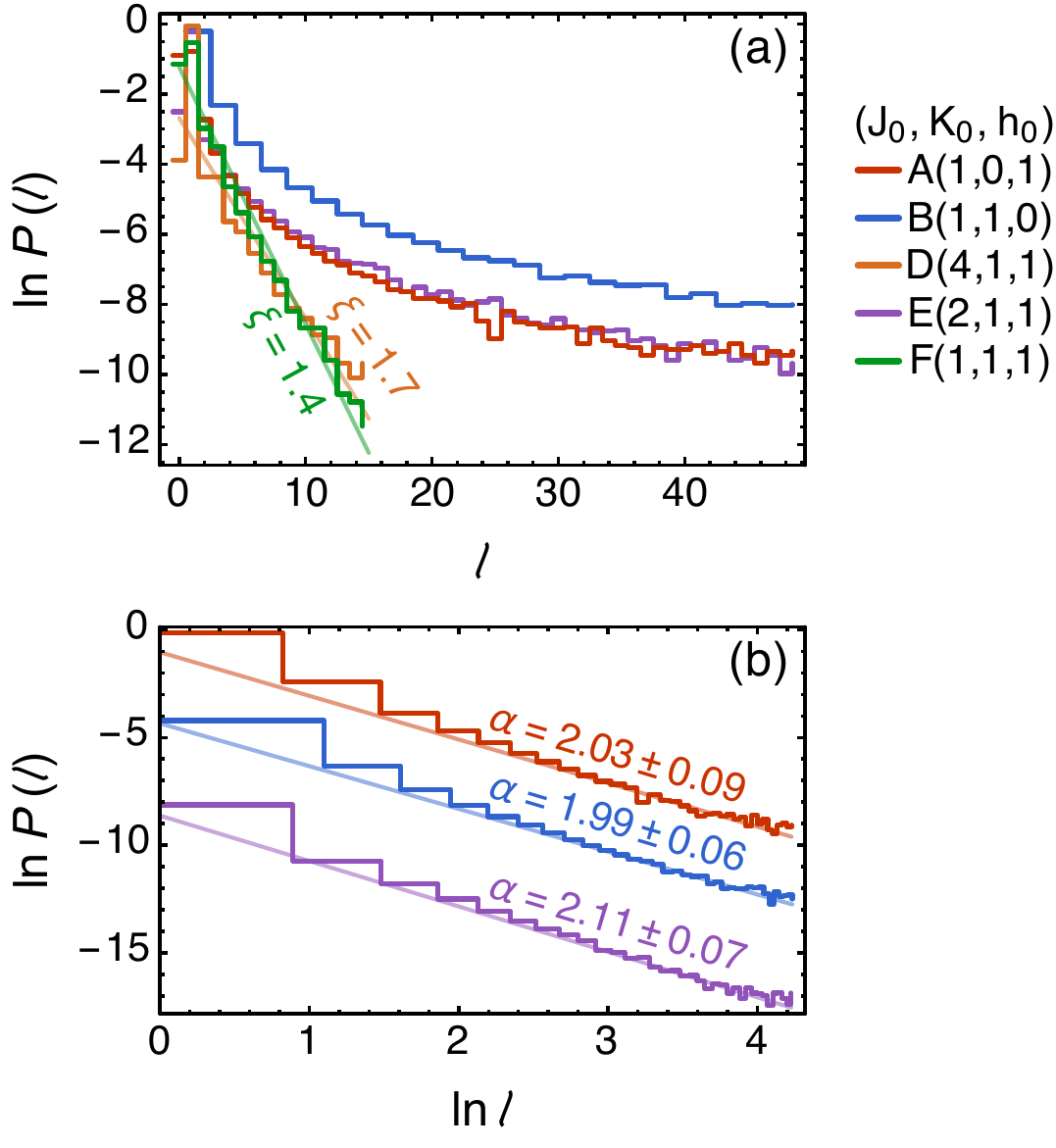}
\caption{Probability distribution $P(l)$ of the stabilizer length $l$, shown as (a) log-linear plot and as (b) log-log plot. The statistics is collected from 1500 random realizations on a 128-site lattice for each parameter $(J_0,K_0,h_0)$ as marked out in the phase diagram \figref{fig: phase diagram} accordingly, with $\Gamma_0=1$. In (b), the lines are shifted vertically from each other (by 4 vertical units) for clarity.}
\label{fig: boundary locality}
\end{center}
\end{figure}

Having the above picture in mind, we now check the stabilizer locality quantitatively. First, we collect the stabilizers obtained by SBRG, and count the number of stabilizers of each length $l$. The length of the stabilizer is defined as the real-space separation from the left-most qubit to the right-most qubit that are acted by the stabilizer in the physical Hilbert space. For each initial scales $(J_0,K_0,h_0)$ of the couplings, we repeat the calculation for many random realizations to obtain reliable statistics. The statistical frequency is then normalized to the probability distribution $P(l)\dd l$ of the stabilizer length $l$. \figref{fig: boundary locality}(a) shows the probability distribution $P(l)$ v.s. $l$ in the log-linear plot. According to the phase diagram \figref{fig: phase diagram} to be shown latter, $D(4,1,1)$ belongs to the SG MBL phase, and $F(1,1,1)$ belongs to the PM MBL phase. We can see, in both MBL phases, the probability distribution decays exponentially $P(l)\sim e^{-l/\xi}$, meaning that the stabilizers are exponentially localized in the MBL phases. Our result can be viewed as an supportive evidence for Ref.\,\onlinecite{Abanin:2013lc,Huse:2014ec} and also agrees with the analysis in Ref.\,\onlinecite{Kim:2014zj}. 

However at the marginal MBL critical point, the stabilizers are no longer short-ranged, but become quasi-long-ranged with the power law distribution $P(l)\sim l^{-\alpha}$. According to the phase diagram in \figref{fig: phase diagram}, $A(1,0,1)$, $B(1,1,0)$ and $E(2,1,1)$ are on (or very close to) the marginal MBL critical line. Then in \figref{fig: boundary locality}(a), their probability distributions $P(l)$ indeed deviate from the exponential decay, and develop long tails for large $l$. To confirm the power law behavior, we switch to the log-log plot in \figref{fig: boundary locality}(b), and find that the data follow straight lines nicely. For the free fermion chains $A(1,0,1)$ and $B(1,1,0)$, the universal exponent $\alpha = 2$ is expected, because according to RSRG fixed point solution,\cite{Fisher:1995cr} the length scale $l$ is related to the logarithmic energy scale $\Gamma =-\ln\epsilon$ as $l\sim \Gamma^2$, so the number of stabilizers identified at that energy scale should follow $P(l)\dd l = (N_0/l)(\dd\Gamma/\Gamma)\sim l^{-2}\dd l$ where $N_0$ is the initial number of sites (qubits). We confirm the exponent $\alpha=2$ in \figref{fig: boundary locality}(b) (see the lines $A$ and $B$). For the interacting fermion chain $E(2,1,1)$ at criticality, the RG fix-point theory is not known to us. Our numerical result seems to imply roughly the same exponent $\alpha\simeq 2.1$ as the line $E$ in \figref{fig: boundary locality}(b). The numerical calculation obtains a slightly larger exponent, because it is difficult to exactly locate the critical line, thus the point $E(2,1,1)$ may be a little off-critical, therefore $P(l)$ is subject to a weak exponential decay, which could result in a seemly larger exponent from the power-law fitting. In \tabref{tab: power}, we collect the exponents $\alpha$ calculated on a few points along (or  near) the marginal MBL critical line. We do not observe systematic or significant deviation from $\alpha = 2$. So our result implies that the lengths of the stabilizers are likely to  follow the same universal scaling behavior $P(l)\sim l^{-2}$ along the whole critical line.

\begin{table}[htdp]
\caption{The scaling exponent $\alpha$ obtained by the power-law fitting along (or near) the phase boundary.}
\begin{center}
\begin{tabular}{cc||cc}
$(J_0,K_0,h_0)$ & $\alpha$ & $(J_0,K_0,h_0)$ & $\alpha$ \\
\hline
$(1,0,1)$ & $2.03\pm0.09$ & $(5,3,2)$ & $2.13\pm0.03$ \\
$(8,1,7)$ & $2.03\pm0.09$ & $(6,4,1)$ & $2.12\pm0.06$ \\
$(4,1,3)$ & $1.98\pm0.09$ & $(7,5,1)$ & $2.06\pm0.06$ \\
$(2,1,1)$ & $2.11\pm0.07$ & $(1,1,0)$ & $1.99\pm0.06$ \\
\end{tabular}
\end{center}
\label{tab: power}
\end{table}%

\begin{figure}[htbp]
\begin{center}
\includegraphics[width=185pt]{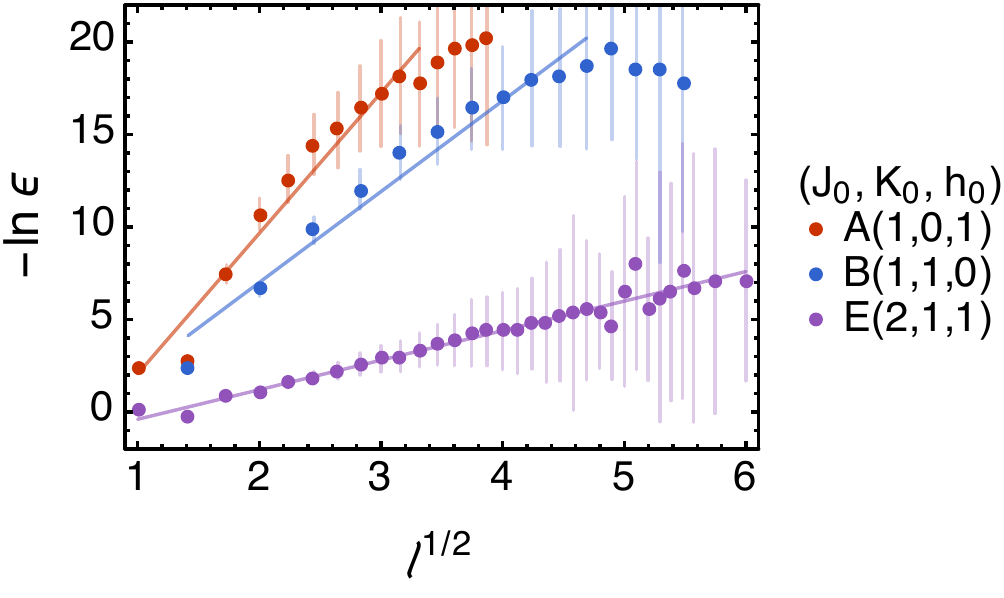}
\caption{Universal dynamical scaling. Different colors represent the different settings of the initial scales $(J_0,K_0,h_0)$ as marked out in the phase diagram \figref{fig: phase diagram} accordingly. The calculation is performed on a 128-site lattice with 1000 random realizations. }
\label{fig: dynamic scaling}
\end{center}
\end{figure}

Another important question is the dynamical scaling at the marginal MBL criticality. The absence of diffusion in the MBL system is characterized by the infinite dynamical exponent  $z\to \infty$ in $l\sim t^{1/z}$. Or more precisely, this implies the logarithmic dynamical scaling $l\sim(\ln t)^\eta$ at the marginal MBL critical point, where $\eta=2$ is expected from the previous RSRG study on the transverse field Ising model.\cite{Fisher:1995cr} We can check the dynamical scaling by studying the energy scale v.s. the length scale of the stabilizers $\hat{\tau}_i$ identified by SBRG. For every $\hat{\tau}_i$, we read off its energy scale $\epsilon_i$ from the effective Hamiltonian $H=\sum_i\epsilon_i\tau_i+\cdots$, and measure its length $l_i$ in the physical space. We collect the pair $(\epsilon_i,l_i)$ for all stabilizers in the system, and repeat the calculation for many random realizations to obtain sufficient amount of samples. \figref{fig: dynamic scaling} shows that the data points roughly follow the $-\ln \epsilon\sim\sqrt{l}$ behavior, and hence the dynamical scaling $l\sim(-\ln \epsilon)^2\sim(\ln t)^2$ is justified. Our data imply that the universal dynamical scaling not only applies to the free fermion chains $A(1,0,1)$, $B(1,1,0)$, but also applies to interacting fermion chain $E(2,1,1)$. Thus we conclude that for the quantum Ising model, the whole marginal MBL critical line is characterized by the same universal dynamical scaling with $z\to \infty$.

\subsection{Entanglement Entropy}

In the Clifford circuit formalism, every energy eigenstate in the physical Hilbert space is approximated by a stabilizer state $\ket{\Psi_{\{\tau_i\}}}$ as in \eqnref{eq: stabilizer state} that corresponds to the direct product state $\ket{\{\tau_i\}}$ in the holographic bulk. In this case, all the entanglement structure of the stabilizer state is given by the Clifford circuit, and the bulk qubits will make no contribution because they are completely disentangled. However, we know the Clifford circuit only provides the 0th order approximation of the eigenstates. If we fix the Clifford circuit, and map the exact physical eigenstate from the holographic boundary into the bulk, then the corresponding bulk state will actually has some entanglement. The entangled bulk qubits will provide an additional layer of entanglement on top of the entanglement given by the Clifford circuit. Therefore the EE $S_E$ of an energy eigenstate is given by the background entropy $S_\text{Cl}$ based on the Clifford circuit  plus the correction $S_\text{bulk}$ due to the bulk qubit entanglement (which can be negative),
\beq
S_E=S_\text{Cl}+S_\text{bulk}.
\eeq
A nice property of the Clifford gates $R_k$ is that they are all of the same MPO bond dimension $D_{R}=2$. So the MPO bond dimension of the Clifford circuit is simply $D=D_{R}^d=2^d$ with $d$ being the circuit depth. This imposes an upper bound on the EE of the Clifford circuit: $S_\text{Cl}\leq \ln D=d\ln 2$. In the MBL phase, the depth of the Clifford circuit is finite, and $S_\text{Cl}$ follows the area-law scaling. At the marginal MBL criticality, the depth of the Clifford circuit grows with the system size logarithmically, so $S_\text{Cl}\sim \ln L$ also follows the logarithmic scaling. Typically the Clifford circuit generated by SBRG can not provide a volume law entanglement scaling, because for an $N$-qubit system, there are only $N$ stabilizers, and most of them are localized. Therefore the volume law entanglement (if any) can only come from the holographic bulk as $S_\text{bulk}\sim L$. But in that case, $S_\text{Cl}$ is overwhelmed by $S_\text{bulk}$ so that the Clifford circuit is no longer a good starting point to encode the volume-law state.

In the strong disorder limit, the bulk contribution $S_\text{bulk}$ can be neglected, and the EE can be estimated from the Clifford circuit alone, i.e. $S_E \simeq S_\text{Cl}$. It turns out that the bipartite EE can be calculated efficiently for stabilizer states,\cite{Chuang:2004ni} and the method is reviewed in Appendix \ref{sec: EE}. For each Clifford circuit generated by SBRG, we can choose a subsystem of the length $L$ and calculate the EE associated to its reduced density matrix using the stabilizer formalism. Then we repeat the calculation with the entanglement cuts translated through out the system to average over the disorder configuration. We also repeat for many random realizations of the system to get more reliable statistics.

\begin{figure}[htbp]
\begin{center}
\includegraphics[width=192pt]{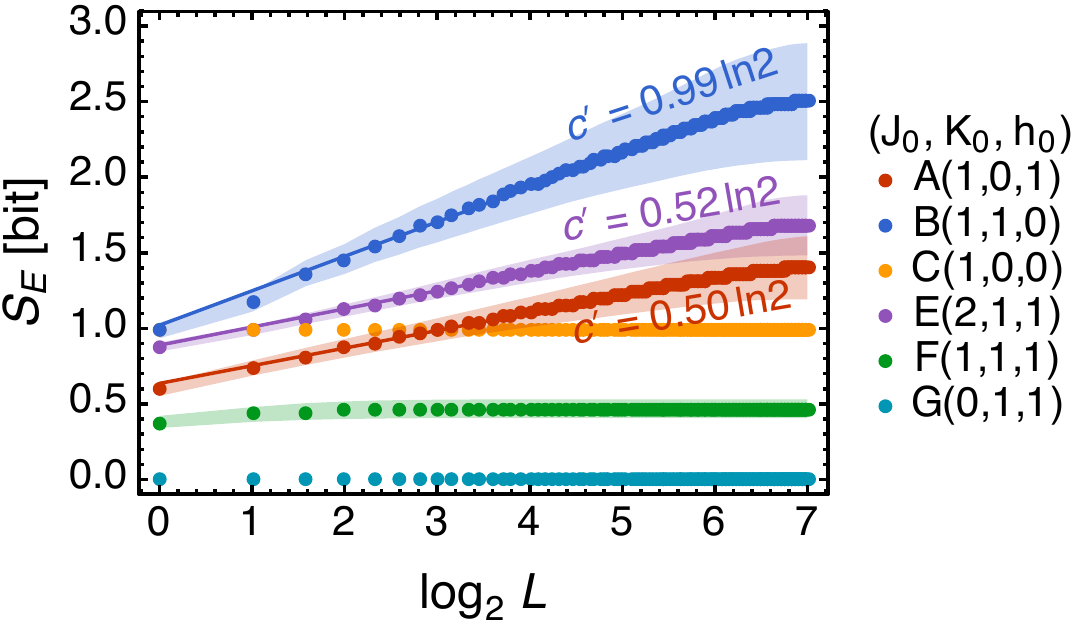}
\caption{Scaling of EE $S_E$ for various initial $(J_0,K_0,h_0)$ with $\Gamma_0=1$. $L$ is the length of the subsystem for the EE measurement. The points are marked out in the phase diagram \figref{fig: phase diagram} correspondingly. The shading denotes the confidence interval of one standard deviation. The calculation is performed on a 256-site lattice with 50 random realizations. The small deviation in the end is a finite size effect when the subsystem size $L$ approaching to the half-system size, i.e. $\log_2 128=7$.}
\label{fig: entropy scaling}
\end{center}
\end{figure}

\figref{fig: entropy scaling} shows our numerical result for the quantum Ising model in \eqnref{eq: Ising int} with coupling coefficients drawn from uniform initial distributions (at $\Gamma_0=1$). At the marginal MBL criticality ($A$, $B$ and $E$), the EE follows the logarithmic scaling:
\beq
S_E(L)=\frac{c'}{3}\ln L.
\eeq
where $c'$ is the effective central charge at the strong disorder fixed-point, which is related to the central charge $c$ of the same system in the clean limit by $c'=c \ln 2$.\cite{Moore:2004ee} The point $A(1,0,1)$ corresponds to a free Majorana chain  with $c=1/2$. The point $B(1,1,0)$ corresponds to two free Majorana chains with $c=1$. The point $E(2,1,1)$ is in between $A$ and $B$, and corresponds to an interacting Majorana chain with $c=1/2$, because $h_i$ is a relevant perturbation which always drives the two Majorana chains into one. The corresponding effective central charges $c'$ in all these cases agree with our numerical results as shown in \figref{fig: entropy scaling}.

In the MBL phase ($C$, $F$ and $G$), the EE saturates to the area law (which is a constant for the 1D system). The point $C(1,0,0)$ is deep in the SG phase, so its EE is exactly 1bit ($S_E=\ln 2$) as expected for the Greenberger-Horne-Zeilinger (GHZ)\cite{Greenberger:2007ty} state of the large block spin (or from the Majorana zero modes at the entanglement cuts). The point $G(0,1,1)$ is deep in the PM phase, so its EE must vanish ($S_E=0$) because all the eigenstates are on-site (atomic) direct product states. The point $F(1,1,1)$ is a generic point of the interacting Majorana chain, whose EE will initially grow a little with $L$ and quickly saturate to the area law at a non-universal value. We can see that all these cases can be handled by SBRG nicely.

\begin{figure}[htbp]
\begin{center}
\includegraphics[width=136pt]{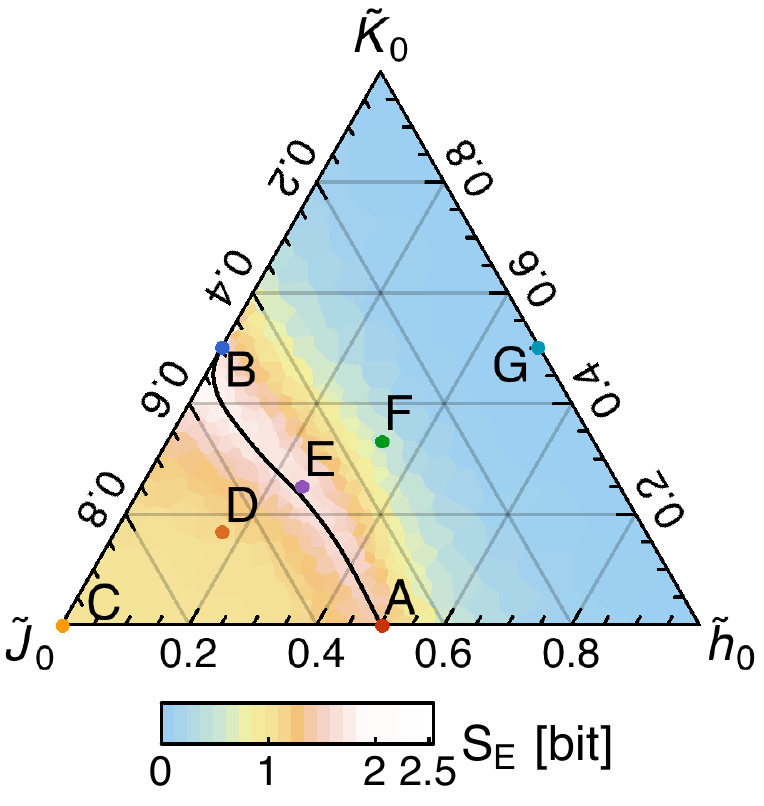}
\caption{Phase diagram of the quantum Ising model \eqnref{eq: Ising int} in ternary graph of $(\tilde{J}_0,\tilde{K}_0,\tilde{h}_0)\equiv(J_0,K_0,h_0)^{1/\Gamma_0}$. The color plot shows the half-system-size EE $S_E$. The black curve traces out the SG-PM phase boundary following the local maximal of $S_E$. The colored points are labeled by their coordinates $(J_0,K_0,h_0)$ as $A(1,0,1)$, $B(1,1,0)$, $C(1,0,0)$, $D(4,1,1)$, $E(2,1,1)$, $F(1,1,1)$, $G(0,1,1)$. The calculation is performed on a 256-site lattice.}
\label{fig: phase diagram}
\end{center}
\end{figure}

Because the EE diverges logarithmically along the phase boundary between the two MBL phases, it can be used to map out the phase diagram. We calculate the half-system-size EE on a sufficiently large system (256-site), and the result is shown in \figref{fig: phase diagram}. The phase boundary is traced out along the local maximum of the EE in the phase diagram. It was also proposed in Ref.\,\onlinecite{Pollmann:2014sf} to detect the phase boundary by the divergent EE fluctuation. We also tried that method as well and obtained basically the same phase diagram (not shown). 

\subsection{Locality of the Effective Hamiltonian}

At the end of SBRG flow, we arrived at the effective Hamiltonian $H_\text{eff}$ of the emergent qubits $\tau_i$ in the holographic bulk.
\beq\label{eq: Heff}
H_\text{eff}=\sum_{i}\epsilon_i\tau_i+\sum_{i<j}\epsilon_{ij}\tau_i\tau_j+\sum_{i<j<k}\epsilon_{ijk}\tau_i\tau_j\tau_k + \cdots,
\eeq
which can be considered as the RG transform of the initial physical Hamiltonian $H$ to the 2nd order precision, as $H_\text{eff}\simeq U_\text{RG}^\dagger H U_\text{RG}$. So how does $H_\text{eff}$ look like in the holographic bulk? Is is still a local Hamiltonian? To answer these questions, we must first assign the position to each emergent qubit in the holographic bulk.

The position of each emergent qubit $\tau_i$ can be labeled by $(x_i,\epsilon_i)$, where $x_i$ is the real-space coordinate of the qubit in the Clifford circuit in \figref{fig: circuit},\footnote{In general $\vect{x}_i$ can be a real-space coordinate vector for higher dimensional problems.} and $\epsilon_i$ is the energy scale associated to $\tau_i$ which is taken from the single-body terms in $H_\text{eff}$. It is reasonable to use the qubit position in the Clifford circuit directly as the real-space coordinate, because the Clifford circuit preserves the real-space locality from the holographic boundary, as it only contains at most $\sim\ln L$ layers of local Clifford gates. If one wishes to be more precise about the real-space coordinate, one can map the emergent qubit back to the stabilizer $\hat{\tau}_{i}$ on the holographic boundary, and use certain center-of-mass coordinate of the stabilizer $\hat{\tau}_i$ as the real-space coordinate for the emergent qubit. However center-of-mass method gives roughly the same coordinate, because the qubit position method already ensures $x_i$ to fall in the real-space support of $\hat{\tau}_i$ and we have demonstrated that $\hat{\tau}_i$ has nice locality in the MBL system (in \figref{fig: boundary locality}).   

\begin{figure}[htbp]
\begin{center}
\includegraphics[width=200pt]{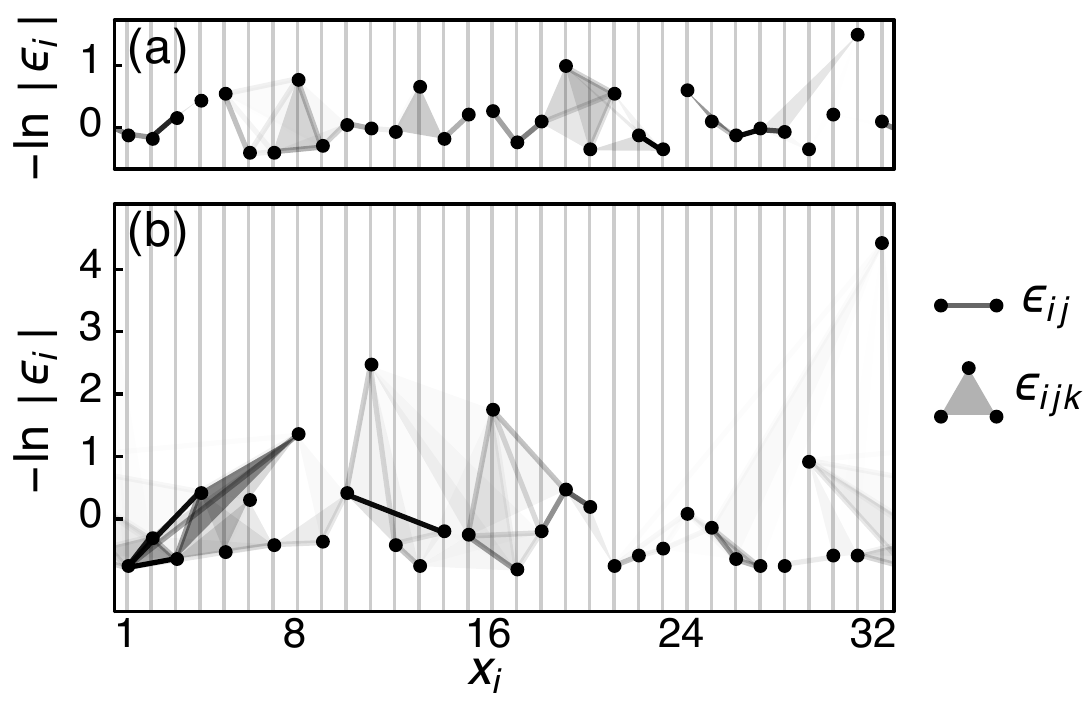}
\caption{Effective Hamiltonian $H_\text{eff}$ in the holographic bulk for (a) $(J_0,K_0,h_0)=(1,1,1)$ and (b) $(J_0,K_0,h_0)=(2,1,1)$ on a 32-site lattice with periodic boundary condition. Each black dot denotes an emergent qubit $\tau_i$ placed in the holographic bulk according to its energy scale $\epsilon_i$. The 2-body (3-body) terms in the effective Hamiltonian are represented by the links (triangles) between the emergent qubits. The darker object corresponds to the stronger interaction term in $H_\text{eff}$.}
\label{fig: Heff}
\end{center}
\end{figure}

As we have pinned down the positions of the emergent qubits $\tau_i$, we can draw pictures of the effective Hamiltonian $H_\text{eff}$ in the holographic bulk. Two typical examples are shown in \figref{fig: Heff}, where the 2-body and 3-body terms are shown respectively as the links and triangles among $\tau_i$. \figref{fig: Heff}(a) shows an example in the MBL phase, which demonstrates nice real-space locality: the links and triangles are typically small, and there is no long object stretching through out the system. \figref{fig: Heff}(b) is an example at the marginal MBL criticality, where the energy scale is extended much deeper into the IR limit, and longer links/triangles are also observed, which is consistent with the delocalization at the criticality. In both pictures, we can see the strong links and strong triangles tend to appear together in the same region, which implies that there should be some correlation of the 2-body terms with the 3-body and higher many-body terms in $H_\text{eff}$, as pointed out in Ref.\,\onlinecite{Lee:2015aa} in a different context.

\begin{figure}[htbp]
\begin{center}
\includegraphics[width=200pt]{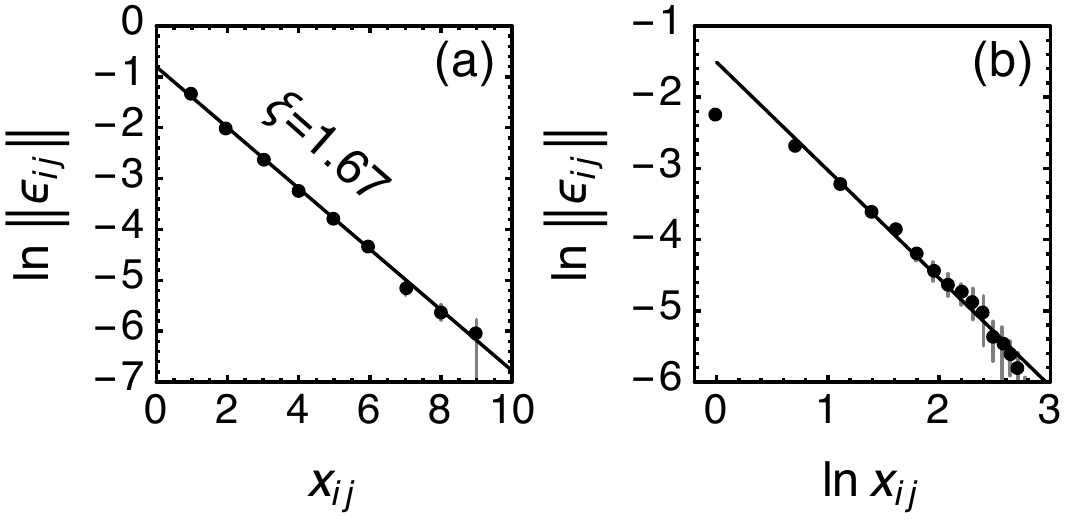}
\caption{Norm of the 2-body energy coefficients $\epsilon_{ij}$ as a function of the distance $x_{ij}$ between the emergent qubits for (a) $(J_0,K_0,h_0)=(1,1,1)$ in log-linear plot and (b) $(J_0,K_0,h_0)=(2,1,1)$ in log-log plot. The calculation is performed on a 32-site lattice with 1000 random realizations.}
\label{fig: Heff locality}
\end{center}
\end{figure}

Now we will focus on the 2-body energy coefficients $\epsilon_{ij}$, and study their locality quantitatively. We classify the $\epsilon_{ij}\tau_i\tau_j$ terms in the effective Hamiltonian $H_\text{eff}$ by their real-space distances $|x_i-x_j|$, and calculate their Frobenius norm as a function of the distance,
\beq
\norm{\epsilon_{ij}}(x_{ij})=\sqrt{\sum_{|x_i-x_j|=x_{ij}}\epsilon_{ij}^2}.
\eeq
It is verified that the norm decays exponentially
\beq \norm{\epsilon_{ij}}\sim e^{-x_{ij}/\xi}\eeq
with the real-space distance $x_{ij}$ in the MBL phase as shown in \figref{fig: Heff locality}(a), which was also proposed in Ref.\,\onlinecite{Abanin:2013lc,Huse:2014ec,Abanin:2015io}. The real-space decay of $\norm{\epsilon_{ij}}$ is related to the logarithmic growth of EE after a global quench in the MBL system,\cite{Moore:2012ge, Huse:2014ec} as $S(t)=s_\infty\xi\ln(J_0 t)$, which is an MBL intrinsic phenomenon that is not present in the Anderson localization. At the marginal MBL criticality, the norm follows a power-law behavior $\norm{\epsilon_{ij}}\sim x_{ij}^{-\alpha}$ as is shown in \figref{fig: Heff locality}(b).

\subsection{Holographic Hamiltonian}

To further study the entanglement structure among the emergent qubits, we map the initial physical Hamiltonian from the holographic boundary into the bulk by the Clifford circuit, and define the bulk Hamiltonian as the \emph{holographic Hamiltonian} $H_\text{hol}$,
\beq
H_\text{hol}=U_\text{Cl}^\dagger H U_\text{Cl}.
\eeq
The holographic Hamiltonian $H_\text{hol}$ acts on the emergent qubits in the holographic bulk. Because the unitary transform $U_\text{Cl}$ can be carried out exactly, $H_\text{hol}$ has exactly the same spectrum as $H$, and its energy eigenstates are the corresponding exact eigenstates of $H$ mapped into the holographic bulk by the Clifford circuit $U_\text{Cl}$.

Because $U_\text{Cl}$ does not exactly diagonalize the Hamiltonian $H$, $H_\text{hol}$ contains off-diagonal terms, which give rise to the entanglement among the emergent qubits in the bulk eigenstates of $H_\text{hol}$. The portion of the off-diagonal terms in the holographic Hamiltonian $H_\text{hol}$ can be measured from the following ratio:
\beq\label{eq: Hhol off-diag}
\frac{\Tr (H_\text{hol} -\text{diag}\,H_\text{hol})^2}{\Tr H_\text{hol}^2}=\frac{\overline{\delta E^2}}{\overline{E^2}},
\eeq
where $\text{diag}\,H_\text{hol}\equiv\sum_{\{\tau_i\}}\ket{\{\tau_i\}}\bra{\{\tau_i\}}H_\text{hol}\ket{\{\tau_i\}}\bra{\{\tau_i\}}$ denotes the diagonal part of $H_\text{hol}$, and on the right-hand-side, $\overline{\delta E^2}$ and $\overline{E^2}$ are respectively the same as \eqnref{eq: delta E2} and \eqnref{eq: E2}. We have checked this ratio $\overline{\delta E^2}/\overline{E^2}$ in \figref{fig: E variance}, and verified that the off-diagonal term does not dominate the holographic Hamiltonian $H_\text{hol}$ for MBL systems. 

To discover the geometry in the holographic bulk, one can in principle diagonalize $H_\text{hol}$ and study the \emph{mutual information} $I_{ij}$ between two emergent qubits $\tau_i$ and $\tau_j$ on the eigenstate,
\beq
I_{ij} = S_{i}+S_{j} - S_{ij},
\eeq
where $S_{i}$ ($S_{j}$) is the EE of each single qubit, and $S_{ij}$ is the EE for both qubits. Then the distance $d_{ij}$ between two qubits can be defined following the proposal in Ref.\,\onlinecite{Qi:2013fm,Qi:2015ct},
\beq
d_{ij} = -\xi \ln\frac{I_{ij}}{I_0},
\eeq
where $I_0=2\ln 2$ is the maximal mutual information between two qubits. In the strong disorder limit, the holographic Hamiltonian $H_\text{hol}$ becomes diagonal, and its diagonal terms also coincide with the effective Hamiltonian $H_\text{eff}$. In this limit, the emergent qubits are disentangled from each other, so they are infinitely far from each other, and the bulk space is fragmented into isolated points.\cite{Lee:2015aa} Away form the strong disorder limit, the off-diagonal terms in $H_\text{hol}$ start to grow (see \figref{fig: E variance}), which leads to the resonance between the emergent conserved quantities and the entanglement among the bulk qubits. From the geometry perspective, the emergent qubits are getting closer to each other, as if the space is contracting.

\begin{figure}[htbp]
\begin{center}
\includegraphics[width=200pt]{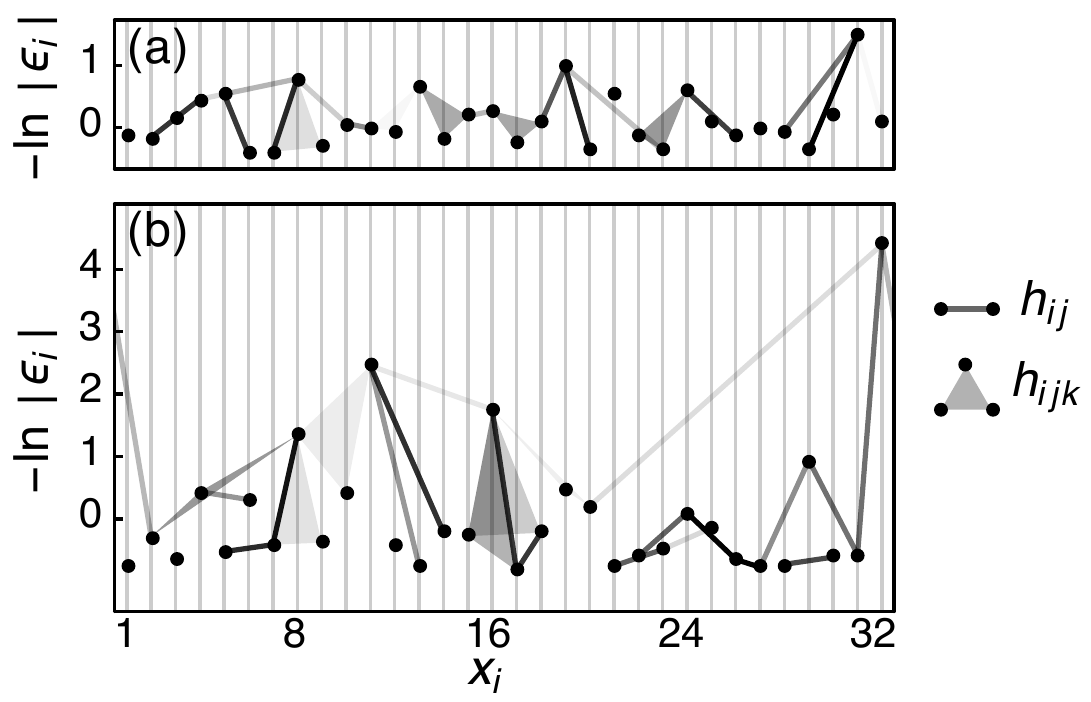}
\caption{Pictures of the off-diagonal terms in the holographic Hamiltonian $H_\text{hol}$ for (a) $(J_0,K_0,h_0)=(1,1,1)$ and (b) $(J_0,K_0,h_0)=(2,1,1)$ on a 32-site lattice with periodic boundary condition. Each black dot denotes an emergent qubit $\tau_i$ in the holographic bulk. The off-diagonal terms of 2-qubit (3-qubit) resonances are represented by the links (triangles) between the emergent qubits. The darker object corresponds to the stronger off-diagonal term in $H_\text{hol}$.}
\label{fig: Hhol}
\end{center}
\end{figure}

To gain more intuition, we draw the off-diagonal terms of $H_\text{hol}$ in the holographic bulk for both the MBL \figref{fig: Hhol}(a) and the marginal MBL \figref{fig: Hhol}(b) systems. The off-diagonal terms create the spacial connectivity in the holographic bulk and drives the emergent qubits closer to each other. In \figref{fig: Hhol}(b) one can also observe a darker resonant cluster\cite{Potter:2015kl} around $x_i\sim16$, in which the qubits are connected by more and stronger off-diagonal terms. We suspect that such regions are closer to thermalization. It is conceivable that if $H_\text{hol}$ is dominated by the off-diagonal terms in a local region of the holographic bulk, the Hamiltonian in that region will look like a random matrix, and the corresponding eigenstates will look like random states,\cite{Page:1993fv} therefore the region is locally thermalized. In the thermalized region, the geometry becomes non-local as every qubit is almost maximally entangled with the rest of the qubits in the region. Therefore the thermalized region can also be viewed as a small blackhole in the holographic bulk.\cite{Maldacena:1999fk,Witten:1998ty} In the MBL-ETH transition, small blackholes will first emerge from the UV region, and then merge into larger blackholes. When a large blackhole covers the IR region, the whole system will be thermalized, since the low-frequency physics is then dominated by the ETH phase.\cite{Altman:2015bh} The above bulk scenario is dual to ergodic puddle percolation on the holographic boundary, which was proposed in Ref.\,\onlinecite{Huse:2014cs,Chandran:2015rc,Potter:2015kl}. 

SBRG approach will break down in the local thermalized region, where the holographic Hamiltonian $H_\text{hol}$ is dominated by strong off-diagonal resonances among the emergent qubits. One key assumption of SBRG is that there exists a set of local quasi-conserved quantities in the MBL system which will not resonate with each other due to their different energy scales, so that the off-diagonal terms can be treated by non-degenerate perturbation. Thus if we found that the off-diagonal terms actually dominate $H_\text{hol}$ after SBRG calculation, it would imply that the assumption fails and the RG result is not reliable. In that case, the emergent qubits found by SBRG in the local thermalized region are no longer quasi-conserved quantities.

Towards thermalization, SBRG becomes not only inaccurate, but also inefficient. Consider a local thermalized region, in which the Hamiltonian $H$ is dominated by the off-diagonal terms, the 2nd order perturbation $H_0+\Sigma\to H_0-\frac{1}{2}\Sigma H_0^{-1}\Sigma$ in \eqnref{eq: H 2nd order} will roughly square the number of terms in $H$ at each RG step. As a result, the number of Hamiltonian terms in the local thermalized region will grow super-exponentially with the RG flow.  The computation complexity of each SBRG step is proportional to the number of Hamiltonian terms, so the RG will quickly get stuck at the boundary of the local thermalized region (also the horizon of the small blackhole) due to the fast growth of the complexity. Although SBRG can not approach the ETH phase, we can still observe indications of local thermalizations in the MBL phase from the increasing off-diagonal resonances in $H_\text{hol}$ or from the growth of the Hamiltonian terms along the RG flow. 

\section{Summary and Discussion}

In summary, we have demonstrated that SBRG is an efficient numerical method to study the MBL (or marginal MBL) systems. The full energy spectrum and all the many-body eigenstates of an MBL system can be approximately found with an algorithm complexity that scales linearly with the system size, which is much more efficient than the ED method which has an exponential complexity. Although we have only explored 1D models in this work as examples, the general formalism of SBRG is readily applicable to higher dimensional models on any lattice. The accuracy of SBRG is controlled by the randomness of the system. We have shown that, deep in the MBL phase, SBRG flows to the strong disorder limit and becomes asymptotically exact. However, as we tune the system towards the ETH phase, SBRG will be more and more inaccurate and inefficient. Although SBRG fails to approach the ETH phase and the thermalization transition, we can still use it to identify small locally thermalized regions in the MBL phase. Nevertheless, the MBL-ETH transition can be described by other RG schemes beyond SBRG, such as the phenomenological RSRG\cite{Huse:2014cs,Potter:2015kl} based on merging the locally thermalized regions. Ref.\,\onlinecite{Chandran:2015rc} further points out that both the MBL and the ETH states can be modeled on the classical level by Clifford circuits. It is interesting to note that the Clifford circuit is naturally generated by SBRG as the classical approximation of the EHM circuit to describe the MBL states. Whether it is possible to design the RG scheme that generates the ergodic Clifford circuit for the ETH state is another interesting problem to explore.

As a Hilbert-space-preserving RG scheme, SBRG can also be interpreted as a realization of the holographic duality. More precisely, an Clifford circuit can be constructed from SBRG flow, which maps the MBL system from the holographic boundary to the bulk. It turns out that the holographic bulk degrees of freedom are the emergent conserved quantities $\tau_i$ of the MBL system which are (approximately) governed by the MBL fixed-point Hamiltonian $H_\text{eff}=\sum_i\epsilon_i\tau_i+\sum_{ij}\epsilon_{ij}\tau_i\tau_j+\cdots$. In the strong disorder limit, the degrees of freedoms in the holographic bulk are disentangled, and the spacial geometry is fragmented. Away from the strong disorder limit, we can show that the locally-thermalized regions emerge in the MBL state, which may be interpreted as the small blackhole formation in the holographic bulk. However, under the present SBRG scheme, the emergent locality along the energy scale can not be observed, and very strong UV-IR mixing prevails in the holographic bulk. We think it is because that each bulk qubit is renormalized independently and abruptly in one RG step, such that the entanglement between the UV and the IR qubits can not be efficiently removed. It will be desirable to improve the RG scheme such that all qubits are renormalized jointly and the Hamiltonian is diagonalized gradually by small unitary transforms.

Note: During the completion of this manuscript, we become aware of a similar work\cite{Rademaker:2015ve} by L. Rademaker, which introduces a different RG scheme that also finds the emergent conserved quantities of the MBL systems by consecutive unitary transforms of the Hamiltonian. The difference is that we keep all orders of interaction terms but treat the off-diagonal resonance perturbatively, while there the resonance is solved exactly but the higher order interactions are truncated.

\begin{acknowledgments}

We would like to acknowledge helpful discussions with Ehud Altman, Guifre Vidal, Sung-Sik Lee, Matthew Fisher, Andrew Potter, Sid  Parameswaran, Bela Bauer, Frank Pollmann, Isaac Kim, Anushya Chandran, Beni Yoshida, Tarun Grover, and Zhi-Cheng Yang. We are especially grateful to Louk Rademaker for sharing with us his unpublished work on the similar topic. YZY and CX are supported by the David and Lucile Packard Foundation and NSF Grant No. DMR-1151208. XLQ is supported by the David and Lucile Packard foundation and NSF Grant No. DMR-1151786. We would also like to acknowledge the hospitality of KITP at Santa Barbara, where the work is initiated during the conference ``Closing the entanglement gap: Quantum information, quantum matter, and quantum fields'', under NSF Grant No. PHY11-25915.

\end{acknowledgments}

\appendix
\section{Technical Details of SBRG}
\subsection{Clifford Group Rotations}\label{sec: R transform}
For our purpose, we generate the Clifford group by the $C_4$ rotation $R_\gate{C4}$ (like $\frac{\pi}{4}$ phase gate) and the swap gate $R_\gate{SWAP}$. The $C_4$ rotation is defined by its generator $\sigma^{[\mu]}$ as
\beq
R_\gate{C4}(\sigma^{[\mu]})\equiv\exp\Big(\frac{\ii\pi}{4}\sigma^{[\mu]}\Big)=\frac{1}{\sqrt{2}}(1+\ii\sigma^{[\mu]}).
\eeq
Its adjoint action on a matrix $\sigma^{[\nu]}$ is given by
\beq\label{eq: C4 rotation}
\begin{split}
\sigma^{[\nu]}\to& R_\gate{C4}^\dagger(\sigma^{[\mu]})\sigma^{[\nu]}R_\gate{C4}(\sigma^{[\mu]})\\
& =
\left\{
\begin{array}{ll}
\sigma^{[\nu]} & \text{if $\sigma^{[\mu]}$, $\sigma^{[\nu]}$ commute,}\\
\ii\sigma^{[\nu]}\sigma^{[\mu]} & \text{if $\sigma^{[\mu]}$, $\sigma^{[\nu]}$ anti-commute.}
\end{array}
\right.
\end{split}
\eeq
The swap gate is specified by $i\leftrightarrow j$, where $i$ and $j$ label the two qubits to be exchanged. For example, on the two-qubit level
\beq
R_\gate{SWAP}(1\leftrightarrow 2)=\frac{1}{2}(\sigma^{00}+\sigma^{11}+\sigma^{22}+\sigma^{33}),
\eeq
which can be generalized to any pair of qubits straightforwardly. The adjoint action of the swap gate $R_\gate{SWAP}(i\leftrightarrow j)$ on a matrix $\sigma^{[\mu]}$ is simply exchanging the indices $\mu_i\leftrightarrow\mu_j$ in $[\mu]$. One can see that both the $C_4$ rotation and the swap gate can be implemented efficiently on the algebraic level by manipulating the Pauli indices of the matrices. In SBRG algorithm, the Hamiltonian is never spelt out as a matrix (not even as a sparse matrix), all we kept is a table of coefficients $h_{[\mu]}$ (the coefficient in front of the matrix $\sigma^{[\mu]}$). The Clifford group rotations is implemented  on the index $[\mu]$ (with a possible sign change of $h_{[\mu]}$ when needed).

According to \eqnref{eq: C4 rotation}, if we want to bring a matrix $\sigma^{[\nu]}$ to another matrix $\sigma^{[\mu]}$ which anti-commutes with the original one, we just need to  perform one $C_4$ rotation $R_\gate{C4}(\ii\sigma^{[\mu]}\sigma^{[\nu]})$:
\beq
\sigma^{[\nu]}\xrightarrow{R_\gate{C4}(\ii\sigma^{[\mu]}\sigma^{[\nu]})}\sigma^{[\mu]}
\eeq
However, if we want to bring a matrix $\sigma^{[\nu]}$ to a commuting matrix $\sigma^{[\mu]}$, we can find an intermediate matrix $\sigma^{[\alpha]}$ which anti-commutes with both $\sigma^{[\nu]}$ and $\sigma^{[\mu]}$, and perform two consecutive $C_4$ rotations, which we called a double-$C_4$ rotation,
\beq
\sigma^{[\nu]}\xrightarrow{R_\gate{C4}(\ii\sigma^{[\alpha]}\sigma^{[\nu]})}\sigma^{[\alpha]}\xrightarrow{R_\gate{C4}(\ii\sigma^{[\mu]}\sigma^{[\alpha]})}\sigma^{[\mu]}.
\eeq
Finally if the certain Pauli indices of a matrix do not appear at the position that we want, we can use the swap gate to permute the indices to the intended position.  

With these, we come up with the following protocol to rotate $\sigma^{[\lambda][\mu]}$ ($\lambda=0,3$ and $\mu=0,1,2,3$) to the block diagonal form $\sigma^{[0\cdots]3[0\cdots]}$:
\begin{itemize}
\item If $[\mu]$ contains the index 1 or 2, such as $\sigma^{[\lambda]1[\nu]}$ or $\sigma^{[\lambda]2[\nu]}$ ($\nu=0,1,2,3$): rotate that index to 3 while eliminating other index to 0 by a $C_4$ rotation as
\beq
\begin{split}
\sigma^{[\lambda]1[\nu]}&\xrightarrow{R_\gate{C4}(-\sigma^{[\lambda]2[\nu]})}\sigma^{[0\cdots]3[0\cdots]},\\
\sigma^{[\lambda]2[\nu]}&\xrightarrow{R_\gate{C4}(\sigma^{[\lambda]1[\nu]})}\sigma^{[0\cdots]3[0\cdots]},
\end{split}
\eeq
then move this index 3 to the intent qubit position by a swap gate if necessary.
\item Else if $[\mu]$ has no index of 1 or 2, but the index 3 exists:
\begin{itemize}
\item If there are multiple indices of 3 in $[\lambda][\mu]$: we need to perform the double-$C_4$ rotation:
\begin{itemize}
\item If there is an index 3 right at the intent qubit position, such as  $\sigma^{[\lambda]3[\kappa]}$ ($\kappa=0,3$): the double-$C_4$ rotation goes as
\beq
\begin{split}
\sigma^{[\lambda]3[\kappa]}
&\xrightarrow{R_\gate{C4}(\sigma^{[\lambda]2[\kappa]})}\sigma^{[0\cdots]1[0\cdots]}\\
&\xrightarrow{R_\gate{C4}(-\sigma^{[0\cdots]2[0\cdots]})}\sigma^{[0\cdots]3[0\cdots]}.
\end{split}
\eeq
\item Else the intent qubit position must host the index 0: then find the last qubit position of index 3, and perform the above double-$C_4$ rotation at that position, and finally use a swap gate to bring that qubit to the intent position.
\end{itemize}
\item Else there is only one index 3 in $[\mu]$: just use a swap gate to bring that qubit to the intent position.
\end{itemize}
\item Else $[\mu]$ is $[0\cdots]$: no Clifford group rotation needed. In fact, such matrix $\sigma^{[\lambda][0\cdots]}$ should have already been ascribed to the effective Hamiltonian and should not actually appear in the residual Hamiltonian.
\end{itemize}

At the end of SBRG flow, the Hamiltonian is diagonalized, and all the emergent conserved quantities are identified as the emergent qubits in the holographic bulk. Each emergent qubit $\tau_i$ is positioned by both its real-space coordinate $x_i$ and its energy scale $\epsilon_i$. Following the above protocol, the emergent qubits $\tau_i$ will be arranged (roughly) by their energy scales $\epsilon_i$, i.e. $\epsilon_1\gtrsim\epsilon_2\gtrsim\cdots$. This has the advantage of improving the algorithm efficiency, because the physical and the emergent Hilbert spaces can be easily separated. However the real-space locality is lost, as the real-space coordinates $x_i$ are scrambled by the swap gates. For many physical applications, it is more desired to reorder the emergent  qubits $\tau_i$ by their real-space coordinates $x_i$. To restore the real-space locality, we simply need to push all the swap gates out of the Clifford circuit, and apply them to the emergent qubits to bring the qubits back to the real-space ordering. When the swap gate passes through a $C_4$ gate, the Pauli indices of the $C_4$ generator should be swapped accordingly. In this paper, we always assume that the real-space locality is restored after SBRG flow, and $\tau_i$ are ordered by their real-space coordinates $x_i=i$. Under this implementation, the Clifford circuit will only contain the $C_4$ rotations and no swap gate.

\subsection{Schrieffer-Wolff Transformations}\label{sec: S transform}

Consider the Hamiltonian $H=H_0+\Delta+\Sigma$, where $H_0=h_3\sigma^{3[0\cdots]}$ is the leading energy scale in the block-diagonal form, $\Delta$ is the rest of the diagonal block
\beq\Delta=\sum_{\lambda=0,3}\sum_{[\mu]}h_{\lambda[\mu]}\sigma^{\lambda[\mu]},\eeq and $\Sigma$ is the off-diagonal block \beq\Sigma=\sum_{\nu=1,2}\sum_{[\mu]}h_{\nu[\mu]}\sigma^{\nu[\mu]}.\eeq The matrices $H_0$, $\Delta$ and $\Sigma$ are all Hermitian, and satisfy
\beq\label{eq: properties of H} H_0^2=h_3^2,\quad H_0\Delta = \Delta H_0,\quad H_0\Sigma = -\Sigma H_0.\eeq 
We can eliminate the off-diagonal block to the 2nd order in $h_{\nu[\mu]}/h_3$ by the Schrieffer-Wolff transformation $H\to S^\dagger H S$, where
\beq
\begin{split}
S&=\exp\Big(-\frac{1}{2h_3^2}H_0\Sigma\Big)\\
&=1-\frac{1}{2h_3^2}H_0\Sigma-\frac{1}{8h_3^2}\Sigma^2+\cdots.
\end{split}
\eeq
Using the properties in \eqnref{eq: properties of H}, it is straightforward to verify that to the 2nd order perturbation we have
\beq
\begin{split}
H&\to S^\dagger HS\\
&\simeq H_0+\Delta+\frac{1}{2h_3^2}H_0\Sigma^2+\frac{1}{2h_3^2}H_0[\Sigma,\Delta].
\end{split}
\eeq
The first three terms are block-diagonal, and the last term is block-off-diagonal, which can be checked by examing their commutation relations with $H_0$. We simply project out the block-off-diagonal part $H_0[\Sigma,\Delta]/(2h_3^2)$. Although it is a 2nd order term, but if we further rotate it to the diagonal block, it will only produce a 4th order correction to the diagonal block, which is negligible. Therefore we end up with the 2nd order effective Hamiltonian in \eqnref{eq: H 2nd order} after the Schrieffer-Wolff transformation.

Since the 2nd order perturbation $H_0\Sigma^2/(2h_3^2)$ will generate many terms and cause the Hamiltonian to grow. So in order to prevent the uncontrolled growth of the Hamiltonian that may crash SBRG program on the computer, we need to control the growth rate by truncating the small terms generated in the 2nd order perturbation. In practice, we first set a maximum growth rate $r$ (say $r=2$). In each RG step, we count the number of terms in $\Sigma$ and denote it as $\mathcal{N}_\Sigma$, then we only keep the leading $r\mathcal{N}_\Sigma$ terms in $H_0\Sigma^2/(2h_3^2)$ and discard the rest of the smaller terms. In this way, the Hamiltonian will not grow too fast, and the complexity of the algorithm is controlled. The truncated terms can be collected for the error estimation afterward. We found that deep in the MBL phase very few terms are truncated, but around the marginal MBL critically more terms will be truncated. One can also adjust the maximum growth rate $r$ to see if SBRG result converges in the $r\to\infty$ limit. We found that in most of the cases, $r=2$ is already good enough to converge the Clifford circuit generated by SBRG.

\section{Entanglement Entropy of Stabilizer States}\label{sec: EE}

Let $\mathfrak{S}=\{\hat{\tau}_i\}$ be the set of emergent conserved quantities represented in the original physical Hilbert space (as Pauli operators). Then every energy eigenstate in the MBL spectrum can be approximated by a state $\ket{\Psi_{\{\tau_i\}}}$ stabilized by $\mathfrak{S}$ as
$\forall i:\hat{\tau}_i\ket{\Psi_{\{\tau_i\}}}=\tau_i\ket{\Psi_{\{\tau_i\}}}$. In the quantum information terminology, $\mathfrak{S}$ is a complete set of stabilizers, and $\ket{\Psi_{\{\tau_i\}}}$ is a stabilizer state.

The bipartite EE of a stabilizer state can be calculated by the highly efficient method developed in Ref.\,\onlinecite{Chuang:2004ni}, which will be briefly reviewed here. Suppose we are interested in the EE $S_A$ in a subsystem $A$ of the state $\ket{\Psi_{\{\tau_i\}}}$:\cite{Horodecki:2009gf}
\beq
\begin{split}
S_A&=-\Tr\rho_A\log_2\rho_A,\\
\rho_A&=\Tr_{\bar{A}} \ket{\Psi_{\{\tau_i\}}}\bra{\Psi_{\{\tau_i\}}},
\end{split}
\eeq
where $\bar{A}$ denotes the complement of $A$. We can first classify the stabilizers $\hat{\tau}_i$ into three sets: the stabilizers only supported in $A$ as $\mathfrak{S}_A=\Tr_{\bar{A}}\mathfrak{S}$, the stabilizers only supported in $\bar{A}$ as $\mathfrak{S}_{\bar{A}}=\Tr_{A}\mathfrak{S}$, and the rest of the stabilizers supported in both $A$ and $\bar{A}$ as $\mathfrak{S}_{A\bar{A}}=\mathfrak{S}-\mathfrak{S}_{A}-\mathfrak{S}_{\bar{A}}$.

The EE is fully determined by $\mathfrak{S}_{A\bar{A}}$. Because $\mathfrak{S}_{A\bar{A}}$ contains all the stabilizers that would be broken by the entanglement cut, and thus can not be used to stabilize the state in $A$ (or in $\bar{A}$), which gives rise to the entropy as the state in the subsystem can not be fully determined. Naively the EE would just be proportional to the number of stabilizers in $\mathfrak{S}_{A\bar{A}}$, however this idea must be refined, because some stabilizers in $\mathfrak{S}_{A\bar{A}}$ are actually hidden stabilizers which can be localized to either $A$ or $\bar{A}$ by multiplication with other stabilizers, and hence not contributing to the EE.

To reveal the hidden stabilizers in $\mathfrak{S}_{A\bar{A}}$, we can rewrite each stabilizer $\hat{\tau}_i\in \mathfrak{S}_{A\bar{A}}$ as a direct product $\hat{\tau}_i = \hat{\tau}_i^{A}\otimes\hat{\tau}_i^{\bar{A}}$ explicitly, where $\hat{\tau}_i^{A(\bar{A})}$ is the part of the Pauli operator $\hat{\tau}_i $ that is supported only in $A$ ($\bar{A}$). Then $\hat{\tau}_i^{A}$ may not be commuting with each other, and the statement is that the hidden stabilizers are in one-to-one correspondence to the center of the Pauli group generated by $\hat{\tau}_i^{A}$. 

To reveal the center, we can first construct the matrix $M$ of anti-commutativity among the operators $\{\hat{\tau}_i^{A}\}$,
\beq
\begin{split}
\forall \hat{\tau}_i,\hat{\tau}_j & \in \mathfrak{S}_{A\bar{A}}:\\
&M_{ij}=\left\{\begin{array}{ll}1 & \text{if }\hat{\tau}_i^{A}\hat{\tau}_j^{A}=-\hat{\tau}_j^{A}\hat{\tau}_i^{A},\\ 0&\text{otherwise.}\end{array}\right.
\end{split}
\eeq
It can be shown that the center corresponds to the null space (the kernel) of the integer matrix $M$ modulo 2. Because the null space basis are hidden stabilizers that will not contribute to the EE, so the nullity must be subtracted, and the EE (in unit of bit) is given by the $\dsZ_2$-rank of $M$
\beq
S_A = \frac{1}{2}\text{rank}\,M \text{ (over $\dsZ_2$)}.
\eeq
The $\dsZ_2$-rank of a matrix can be found using Gaussian elimination, and the EE can be simply calculated from the algebraic structure of the stabilizers.

\bibliography{SBRG}
\bibliographystyle{apsrev}

\end{document}